\documentclass[10pt,journal,compsoc]{IEEEtran}
\usepackage{graphicx}
\usepackage{amsmath,amssymb}
\usepackage{color}
\usepackage{overpic}
\usepackage[ruled]{algorithm2e}
\usepackage{soul}
\usepackage{xcolor,colortbl}
\usepackage{multirow}
\usepackage{url}
\usepackage{booktabs}
\usepackage{bigstrut}
\usepackage{subcaption}
\usepackage{lipsum}
\usepackage{multicol}
\usepackage{diagbox}
\usepackage{array}
\newcolumntype{P}[1]{>{\centering\arraybackslash}p{#1}}
\newcolumntype{M}[1]{>{\centering\arraybackslash}m{#1}}

\definecolor{darkcyan}{rgb}{0.0, 0.55, 0.55}

\DeclareMathOperator*{\argmin}{arg\,min}

\ifCLASSOPTIONcompsoc
  \usepackage[nocompress]{cite}
\else
  \usepackage{cite}
\fi

\newcommand{\BLK}{\textcolor{white}}

\ifCLASSINFOpdf

\else

\fi

\begin{document}

\title{Learning End-to-End Lossy Image Compression: A Benchmark}

\author{
		Yueyu~Hu,~\IEEEmembership{Graduate Student Member,~IEEE,}
		Wenhan~Yang,~\IEEEmembership{Member,~IEEE,}\\
		Zhan Ma,~\IEEEmembership{Senior Member,~IEEE,}
		and~Jiaying~Liu,~\IEEEmembership{Senior Member,~IEEE}
		\IEEEcompsocitemizethanks{
			\IEEEcompsocthanksitem Yueyu Hu, Wenhan Yang, and Jiaying Liu are with Wangxuan Institute of Computer Technology, Peking University, Beijing, 100080, China, E-mail: \{huyy, yangwenhan, liujiaying\}@pku.edu.cn.
			\IEEEcompsocthanksitem Zhan Ma is with Electronic Science and Engineering School, Nanjing University, Jiangsu, 210093, China, E-mail: mazhan@nju.edu.cn.
		}
		\thanks{
		(Corresponding author: Jiaying Liu)
		}
		}

\markboth{IEEE Transactions on Pattern Analysis and Machine Intelligence}%
{Shell \MakeLowercase{\textit{et al.}}}

\IEEEtitleabstractindextext{
\begin{abstract}
Image compression is one of the most fundamental techniques and commonly used applications in the image and video processing field. Earlier methods built a well-designed pipeline, and efforts were made to improve all modules of the pipeline by handcrafted tuning. Later, tremendous contributions were made, especially when data-driven methods revitalized the domain with their excellent modeling capacities and flexibility in incorporating newly designed modules and constraints. Despite great progress, a systematic benchmark and comprehensive analysis of end-to-end learned image compression methods are lacking. In this paper, we first conduct a comprehensive literature survey of learned image compression methods. The literature is organized based on several aspects to jointly optimize the rate-distortion performance with a neural network, \textit{i.e.}, network architecture, entropy model and rate control. We describe milestones in cutting-edge learned image-compression methods, review a broad range of existing works, and provide insights into their historical development routes. With this survey, the main challenges of image compression methods are revealed, along with opportunities to address the related issues with recent advanced learning methods. This analysis provides an opportunity to take a further step towards higher-efficiency image compression. By introducing a coarse-to-fine hyperprior model for entropy estimation and signal reconstruction, we achieve improved rate-distortion performance, especially on high-resolution images. Extensive benchmark experiments demonstrate the superiority of our model in rate-distortion performance and time complexity on multi-core CPUs and GPUs. Our project website is available at \url{https://huzi96.github.io/compression-bench.html}.
\end{abstract}

}

\maketitle

\IEEEdisplaynontitleabstractindextext
\IEEEpeerreviewmaketitle

\IEEEraisesectionheading{\section{Introduction}\label{sec:introduction}}

\IEEEPARstart{I}mage compression is a fundamental technique in the signal processing and computer vision fields.
The constantly developing image and video compression methods facilitate the continual innovation of new applications, \textit{e.g.} high-resolution video streaming and augmented reality. The goal of image compression, especially lossy image compression, is to preserve the critical visual information of the image signal while reducing the bit-rate used to encode the image for efficient transmission and storage. For different application scenarios, trade-offs are made to balance the quality of the compressed image and the bit-rate of the code.

In recent decades, a variety of codecs have been developed to optimize the reconstruction quality with bit-rate constraints. In the design of existing image compression frameworks, there are two basic principles. First, the image signal should be decorrelated, which is beneficial in improving the efficiency of entropy coding. Second, for lossy compression, the neglected information
should have the least influence on the reconstruction quality,
\textit{i.e.}, only the least important information for visual experience is discarded in the coding process.

The traditional transform image compression pipeline consists of several basic modules, \textit{i.e.} transform, quantization and entropy coding. A well-designed transform for image compression transforms the image signal into compact and decorrelated coefficients. Discrete Cosine Transform (DCT) is applied to the $8\times8$ partitioned images in the JPEG~\cite{marcellin2000overview} image coding standard. Discrete Wavelet Transform (DWT) in JPEG 2000~\cite{rabbani2002overview} further improves coding performance by introducing a multiresolution image representation to decorrelate images across scales.
Then, quantization discards the least significant information by truncating less informative dimensions in the coefficient vectors. Methods are introduced to improve quantization performance, including vector quantization~\cite{gersho2012vector}
and trellis-coded quantization~\cite{marcellin1990trellis}.
After that, the decorrelated coefficients are compressed with entropy coding.
Huffman coding is first employed in JPEG images.
Then, improved entropy coding methods such as arithmetic coding~\cite{witten1987arithmetic} and context-adaptive binary arithmetic coding~\cite{marpe2003context} are utilized in image and video codecs~\cite{sullivan2012overview}.
In addition to these basic components, modern video codecs, \textit{e.g.}
HEVC and VVC~\cite{vtm8}, employ intra prediction and an in-loop filter for intra-frame coding. These two components are also applied to BPG~\cite{bpg}, an image codec, to further reduce spatial redundancy and improve the quality of the reconstruction frames, especially interblock redundancy. However, the widely used traditional hybrid image codecs have limitations. First, these methods are all based on partitioned blocks of images, which introduce blocking effects.
Second, each module of the codec has a complex dependency on others. Thus, it is difficult to jointly optimize the whole codec. Third, as the model cannot be optimized as a whole, the partial improvement of one module may not bring a gain in overall performance, making it difficult to further improve the sophisticated framework.

With the rapid development of deep learning, there have been many works exploring the potential of artificial neural networks to form an end-to-end optimized image compression framework. The development of these learning-based methods has significant differences from traditional methods. For traditional methods, improved performance mainly comes from designing more complex tools for each component in the coding loop. Deeper analysis can be conducted on the input image, and more adaptive operations can be applied, resulting in more compact codes. However, in some cases, although the performance of the single module is improved, the final performance of the codec, \textit{i.e.}, the superimposed performance of different modules, might not increase much, making further improvement difficult. For end-to-end learned methods, as the whole framework can be jointly optimized, performance improvement in a module naturally leads to a boost to the final objective. Furthermore, joint optimization causes all modules to work more adaptively with each other.

In the design of an end-to-end learned image compression method, two aspects are considered. First, if the latent representation coefficients after the transform network are less correlated, more bit-rate can be saved in the entropy coding.
Second, if the probability distribution of the coefficients can be accurately estimated by an entropy model, the bit-stream can be utilized more efficiently and the bit-rate to encode the latent representations can be better controlled, thus, a better trade-off between the bit-rate and the distortion can be achieved.
The pioneering work of Toderici \textit{et al.}~\cite{toderici2015variable} presents an end-to-end learned image compression that reconstructs the image by applying a recurrent neural network~(RNN).
Meanwhile, generalized divisive normalization~(GDN)~\cite{balle2015density} was proposed by Ball{\'e} \textit{et al.} to model image content with a density model, which shows an impressive capacity for image compression.
Since that time, there have been numerous end-to-end learned image compression methods inspired by these frameworks.

Although tremendous progress has been made in end-to-end learned image compression, there is a lack of a systematic survey and benchmark to summarize and compare different methods thoroughly.
To this end, in this work, we conduct a comprehensive survey of recent progress in learning-based image compression as well as a thorough benchmarking analysis on different methods of learning-based image compression. The contributions and novelties of existing works are summarized and highlighted, and future directions are illustrated.
With the summarized guidance from the survey and benchmark, we propose a novel end-to-end learned image compression framework that offers state-of-the-art performance.

The contributions of this paper are as follows:

\begin{itemize}
\item
We comprehensively summarize the existing end-to-end learned image compression methods.
The contributions and novelties of these methods are discussed and highlighted.
The technical improvements of these methods are commented on based on their categorizations, and we demonstrate a clear picture of the design methodologies and shows interesting future research directions.

\item Inspired by the insights and challenges summarized for the existing approaches, we further explore the potential of end-to-end learned image compression and propose a coarse-to-fine hyperprior modeling framework for lossy image compression. The proposed method is shown to outperform existing methods in terms of coding performance, while keeping the time complexity low on parallel computing hardwares.

\item We conduct a thorough benchmark analysis to compare the performance of existing end-to-end compression methods, the proposed method, and traditional codecs. The comparison is conducted fairly from different perspectives, \textit{i.e.} the rate-distortion performance on different ranges of bit-rate or resolution and the complexity of the implementation.
\end{itemize}

Note that, this paper is the extension of our earlier publication~\cite{hu2020coarse}. We summarize the changes here. First, this paper additionally focuses on the thorough survey and benchmark of end-to-end learned image compression methods. In addition to \cite{hu2020coarse}, we summarize the contributions of existing works on end-to-end learned image compression in Sec.~\ref{sec:overview}, and present a more detailed comparative analysis of the merits towards high-efficiency end-to-end learned image compression in Sec.~\ref{sec:backbone} and Sec.~\ref{sec:entropy}. Second, we conduct a benchmark evaluation of existing methods in Sec.~\ref{sec:rdperformance}, where we present the comparative experimental results on two additional datasets, in both PSNR and MS-SSIM. Third, we raise the novel problem of cross-metric performance with respect to image compression methods in Sec.~\ref{sec:crossmetric}, where we present the empirical analysis on the phenomenon of cross-metric bias and we briefly discuss future research directions to address the related issues.

The rest of the paper is organized as follows. In Sec.~\ref{sec:formulation}, we first formulate the image compression problem, especially focusing on end-to-end learned schemes. After that, in Sec.~\ref{sec:overview}, we briefly summarize the main contributions of existing research. Then, in Sec.~\ref{sec:backbone}, we categorize existing learned image compression methods according to their backbone models. After that, special attention is paid to the rate control technique in Sec.~\ref{sec:entropy}, which is the very specialized component in image compression compared with other deep-learning processing or understanding methods. Inspired by our survey and analysis, we introduce our new proposed method in Sec.~\ref{sec:propose}. Later, in Sec.~\ref{sec:evaluation}, we introduce the benchmarking protocols and make benchmarking comparisons of existing methods. Finally, in Sec.~\ref{sec:conclude}, we draw conclusions and discuss potential future research directions.

\section{Problem Formulation}
\label{sec:formulation}

Natural image signals include many spatial redundancies and have the potential to be compressed without much degradation in perceptual quality. Considering practical constraints on bandwidth and storage, lossy image compression is widely adopted to minimize the bit-rate of representing a given image to tolerate a certain level of distortion.
The compression framework usually consists of an encoder-decoder pair. Given an input image $\mathbf{x}$ with its distribution $p_{\mathbf{x}}$, the encoder with an encoding transform $\mathcal{E}$ and a quantization function $\mathcal{Q}$, a discrete code $\mathbf{y}$ is generated as follows:
\begin{equation}
	\mathbf{y} = \mathcal{Q}(\mathcal{E}(\mathbf{x}; \theta_{\mathcal{E}})),
\end{equation}
where $\theta_\mathcal{E}$ denotes the encoder parameters to be tuned during the learning procedure. To obtain the pixel representation of the image, the corresponding decoder $\mathcal{D}$ reconstructs the image $\hat{\mathbf{x}}$ from the code $\mathbf{y}$ as follows:
\begin{equation}
	\hat{\mathbf{x}} = \mathcal{D}(\mathbf{y}; \theta_\mathcal{D}) = \mathcal{D}(\mathcal{Q}(\mathcal{E}(\mathbf{x}; \theta_{\mathcal{E}})); \theta_\mathcal{D}),
\end{equation}
where $\theta_\mathcal{D}$ denotes the parameters in $\mathcal{D}$.

Two kinds of metrics, \textit{i.e.} distortion $D$ and bit-rate $R$,
give rise to rate-distortion optimization $R + \lambda D$, the core problem of lossy image compression. The distortion term $D$ measures how different the reconstructed image is from the original image, and it is usually measured via fidelity-driven metrics or perceptual metrics as follows:
\begin{equation}
D = \mathbb{E}_{\mathbf{x} \sim p_{\mathbf{x}}} [d(\mathbf{x}, \hat{\mathbf{x}})],
\end{equation}
where $d$ denotes the distortion function. The rate term $R$ corresponds to the number of bits to encode $\mathbf{y}$, which is bounded according to the entropy constraints. However, the actual probability distribution of the latent code $\mathbf{y}$, denoted as $p_{\mathbf{y}}$, is unknown, making accurate entropy calculation intractable. Thus, we usually utilize an entropy model $q_{\mathbf{y}}$ to serve as the estimation of $p_{\mathbf{y}}$ for entropy coding. Hence, the rate term can be formulated as the cross entropy of $p_{\mathbf{y}}$ and $q_{\mathbf{y}}$ as follows:
\begin{equation}
	R = H(p_{\mathbf{y}}, q_{\mathbf{y}}) = \mathbb{E}_{\mathbf{y} \sim p_{\mathbf{y}}}[-\log q_{\mathbf{y}}(\mathbf{y})],
\end{equation}
where $p_{\mathbf{y}}$ stands for the real probability distribution and $q_{\mathbf{y}}$ refers to the distribution estimated by the entropy model. The overall compression model can be viewed as an optimization of the weighted sum of $R$ and $D$. Formally, the problem can be solved by minimizing the following optimization with a trade-off coefficient $\lambda$ as follows:
\begin{equation}
	\label{eq:rdo}
	\hat{\theta}_E, \hat{\theta}_D, \hat{\theta}_p = \argmin_{\theta_E, \theta_D, \theta_p} R + \lambda D,
\end{equation}
where $\theta_p$ denotes the parameter for the entropy model. The optimal parameters $\hat{\theta}_E, \hat{\theta}_D, \hat{\theta}_p$ cause the model to achieve an overall good rate-distortion performance on the image $\mathbf{x}$ that follows $\mathbf{x} \sim p_{\mathbf{x}}$. Different $\lambda$ values indicate different rate-distortion trade-offs, depending on the requirements of different applications.

Though the idea of rate-distortion optimization is also applied to traditional compression schemes, learning-based methods finally make the joint optimization of all the components feasible. The opportunities and challenges are listed below:
\begin{itemize}
\item \textbf{Global Optimization.} The major difference between learned image compression and the traditional hybrid codec lies in their optimizations. Instead of hand-craft tuning, learned image compression models can be automatically tuned to any differentiable metric, \textit{e.g.} SSIM~\cite{wang2004image}, MS-SSIM~\cite{wang2003multiscale} and perceptual difference~\cite{johnson2016perceptual}, which is calculated by neural networks. In addition, while the traditional hybrid coding framework is usually improved at the scale of individual components, in learning-based methods, all modules are trainable, and it is possible to optimize all parameters and components jointly. However, it is nontrivial to acquire good performance in end-to-end learning compression because of the difficulties in optimization.
\item \textbf{Full-Resolution Processing.} Convolutional neural networks support the full-resolution processing of images, while hybrid frameworks usually process partitioned blocks. Full processing can bring more benefits to entropy modeling with more context and avoid the blocking effect caused by partitioning. Full-resolution processing also comes with an increase in complexity. Because the perceptive field of a convolutional kernel is limited, the network needs to be deepened to perceive more large regions and improve modeling capacity.
\item \textbf{Rate Control.} With joint optimization, the whole model can directly target the rate-distortion constraint, while in hybrid schemes, the additional rate-control component is employed and may not produce an optimal approximation. However, for a large portion of learning-based methods, multiple models need to be trained for different rate-distortion trade-offs. The other single-model variable-bit-rate architectures are usually much more time-consuming. Therefore, practical applications of these methods are sometimes limited.
\end{itemize}

\begin{table*}[htbp]
	\caption{Summary of important contributions of image compression in recent years.}
	\label{tab:sum}
	\centering
\begin{tabular}{M{7.8em}M{18em}cm{27.0em}}

	\toprule
	Method Name & Paper Title & Published In & \multicolumn{1}{M{27.585em}}{Highlight}\\
	\midrule
	Variable-Rate RNN~\cite{toderici2015variable} & Variable Rate Image Compression with Recurrent Neural Networks & ICLR 16 & The first work to utilize a convolutional LSTM network for variable-bit-rate end-to-end learned image compression. \\
	\midrule
	GDN Transform~\cite{balle2016optimization} & End-to-End Optimization of Nonlinear Transform Codes for Perceptual Quality & PCS 16 &  Introduces GDN, a trainable decorrelation nonlinear normalization that shows a great capability for image compression. \\
	\midrule
	Full-Resolution RNN~\cite{toderici2017full} & Full Resolution Image Compression with Recurrent Neural Networks & CVPR 17 &   The first practical recurrent model for variable-bit-rate full-resolution image compression.\\
	\midrule
	Soft-to-Hard Quantization~\cite{agustsson2017soft} & {Soft-to-Hard Vector Quantization for End-to-End Learning Compressible Representations} & NeurIPS 17 & Introduces vector quantization for learned compression and proposes using soft-to-hard annealing techniques to improve the performance of networks with quantization. \\
	\midrule
	Compressive Autoencoder~\cite{theis2017lossy} & Lossy Image Compression with Compressive Autoencoder & ICLR 17 & Residual network is first employed for CNN-based image compression models. A Laplace-smoothed histogram is used as the entropy model. \\
	\midrule
	GDN Network~\cite{balle2016end} & End-to-End Optimized Image Compression & ICLR 17 & Introduces the multilayer nonpartitioning end-to-end architecture with GDN for image compression. \\
	\midrule
	Inpainting Based~\cite{baig2017learning} & Learning to Inpaint for Image Compression & NeurIPS 17 & Utilizes image inpainting techniques in a recurrent framework to improve compression performance. \\
	\midrule
	Real-Time Adversarial~\cite{rippel2017real} & Real-Time Adaptive Image Compression & ICML 17 & The first method to adopt a multiscale framework with adversarial loss for learned real-time image compression.\\
	\midrule
	Tiled Network~\cite{minnen2017spatially} & {Spatially Adaptive Image Compression Using a Tiled Deep Network} & ICIP 17 & Introduces explicit intraprediction with a tiled structure in the network. \\
	\midrule
	Hyperprior~\cite{balle2018variational} & Variational Image Compression with a Scale Hyperprior & ICLR 18 & The first work to propose a hyperprior for image compression, which greatly advances the compression performance. \\
	\midrule
	Context Model~\cite{minnen2018joint} & Joint Autoregressive and Hierarchical Priors for Learned Image Compression & NeurIPS 18 & Proposes combining the spatial context-model and a hyperprior for conditional entropy estimation. \\
	\midrule
	Local Entropy Model~\cite{minnen2018image} & Image-Dependent Local Entropy Models for Learned Image Compression & ICIP 18 & Aims to better encode latent representations with an offline dictionary. \\
	\midrule
	3D-CNN Entropy Model~\cite{mentzer2018conditional} & Conditional Probability Models for Deep Image Compression & CVPR 18 & 3D-CNN is used for learning a conditional probability model for a multiresidual-block-based network. \\
	\midrule
	Priming RNN~\cite{johnston2018improved} & Improved Lossy Image Compression with Priming and Spatially Adaptive Bit Rates for Recurrent Networks & CVPR 18 & The recurrent compression model is improved with a proposed priming technique and spatial contextual entropy model. \\
	\midrule
	Content-Weighted~\cite{li2018learning} & Learning Convolutional Networks for Content-Weighted Image Compression & CVPR 18 & Proposes using a learned importance map to guide the allocation of bits for latent code. \\
	\midrule
	Generative Model~\cite{tschannen2018deep} & Deep Generative Models for Distribution-Preserving Lossy Compression & NeurIPS 18 & GAN is first used for extremely low bit-rate image compression. \\
	\midrule
	Multiscale CNN~\cite{nakanishi2018neural} & Neural Multi-Scale Image Compression & ACCV 18 & Proposes a multiscale model and corresponding contextual entropy estimation to improve compression efficiency. \\
	\midrule
	Intraprediction in Codes~\cite{klopp2018learning} & Learning a Code-Space Predictor by Exploiting Intra-Image-Dependencies & BMVC 18 & Explicitly designs code-space intraprediction to reduce coding redundancy. \\
	\midrule
	Nonuniform Quantization~\cite{cai2018deep} & Deep Image Compression with Iterative Non-Uniform Quantization & ICIP 18 & Proposes nonuniform quantization to reduce quantization error in the network. \\
	\midrule
	Context Model~\cite{lee2018context} & Context Adaptive Entropy Model for End-To-End Optimized Image Compression & ICLR 19 & Introduces a different approach to combine a hyperprior and the context model for image compression. \\
	\midrule
	Energy Compaction~\cite{cheng2019learning} & Learning Image and Video Compression through Spatial-Temporal Energy Compaction & CVPR 19 & Introduces a subband coding energy compaction technique for CNN-based image compression. \\
	\midrule
	GMM \& Attention~\cite{cheng2020learned} & Learned Image Compression with Discretized Gaussian Mixture Likelihoods and Attention Modules & CVPR 20 & 
	Utilizes Gaussian Mixture Model to estimate likelihoods of symbols more accurately. Attention modules are included for improved transform capability. \\
	\midrule
	iWave++~\cite{ma2020end} & End-to-End Optimized Versatile Image Compression with Wavelet-Like Transform & TPAMI 20 & Adopts lifting to build the wavelet-like transforms with neural networks. It simultaneously supports lossy and lossless image compression.\\
	\midrule
	Non-Local \& 3D-Context~\cite{chen2019neural} & Neural Image Compression via Non-Local Attention Optimization and Improved Context Modeling & TIP 21 &Utilizes non-local network and 3D context model to achieve improved rate-distortion performance. \\
	\bottomrule
	\end{tabular}%
\end{table*}

\section{Overview of Progress in Recent Years}
\label{sec:overview}
Since the pioneering work of Toderici \textit{et al.}~\cite{toderici2015variable} in 2015 exploited recurrent neural networks for learned image compression, much progress has been made.
Benefiting from the strong modeling capacity of deep networks, the performance of learned image compression has exceeded that of JPEG to BPG (HEVC Intra), and the performance gap is widening further.
The milestones of learned image compression are summarized in Table~\ref{tab:sum}. Early works aim to search for possible architectures to apply transform coding with neural networks and propose end-to-end trainable solutions.
Ball{\'e}~\textit{et al.}~\cite{balle2015density,balle2016optimization,balle2016end} proposes a learning-based framework with GDN nonlinearity embedded analysis and synthesis transforms for learned image compression, while Toderici ~\textit{et al.} utilize recurrent models for variable-rate learned compression~\cite{toderici2015variable,toderici2017full}.

To make the network end-to-end trainable,
the quantization component, which is not differentiable based on the definition, should be designed carefully and approximated by a differentiable process.
Some works replace the true quantization with additive uniform noise~\cite{balle2016end,balle2018variational}
while others use direct rounding in forwarding and back-propagate the gradient of $y=x$.
In addition, Agustsson \textit{et al.}~\cite{agustsson2017soft} proposes replacing direct scalar quantization with soft-to-hard vector quantization to make the quantization smoother. Dumas \textit{et al.}~\cite{dumas2018autoencoder} designs a model that additionally learns the quantization parameters. As it is non-trivial to train a variational autoencoder~(VAE)~\cite{kingma2013auto} based model that incorporates quantization, advanced optimization techniques for image compression are still being extensively studied recently~\cite{yang2020improving}.

When the compression network is trainable, the next issue is to efficiently reduce spatial redundancy in the image signal,
where the transform is usually a critical part.
Some take the form of a convolutional neural network (CNN), \textit{e.g.} GDN~\cite{balle2015density,balle2016end,balle2016optimization} or residual block with enhanced nonlinearity~\cite{theis2017lossy}. Some advanced convolutional architectures like attention module~\cite{cheng2020learned}, non-local networks~\cite{chen2019neural}, and invertible structures~\cite{ma2020end} have also been employed to improve the modeling capacity of the transforms.
Others resort to a recurrent neural network (RNN) to infer latent representations progressively, which forms a scalable coding framework~\cite{toderici2015variable}. In each iteration, the network largely squeezes out the unnecessary bits in the latent representations. Therefore, the final representations are compact.

After the transform, the compact latent representations are further compressed via entropy coding, where frequently occurring patterns are represented with few bits and rarely occurring patterns with many bits.
Earlier works incorporate elementwise independent entropy models to estimate the probability distribution of the latent representations~\cite{balle2016end,theis2017lossy} and independently encode each element with an arithmetic coder.
With these initial trials, later advanced methods explicitly estimate entropy with hyperpriors~\cite{balle2018variational,minnen2018image},
predictive models~\cite{minnen2018joint,lee2018context,klopp2018learning,cheng2020learned,ma2020end,chen2019neural} or
other learned parametric models~\cite{toderici2017full,johnston2018improved,mentzer2018conditional}.

In addition to the abovementioned methods that target signal fidelity with learned transform coding frameworks, there are emerging works targeting novel application conditions, notably compression for machine vision~\cite{duan2020video} or human perception at low bit-rates. 
According to research on the human visual system, human eyes are less sensitive to pixelwise distortion in areas with complex texture. Therefore, generative models such as conditional generative adversarial networks (GAN) can be employed to synthesize such areas, where low-bit-rate representations can serve as the guidance. This can be utilized to design high-efficiency image codecs.
Rippel \textit{et al.}~\cite{rippel2017real} first proposed utilizing the adversarial loss function in an end-to-end framework to improve visual quality. In later literature, Agustsson \textit{et al.}~\cite{agustsson2018generative}, Tschannen \textit{et al.}~\cite{tschannen2018deep} and Santurkar \textit{et al.}~\cite{santurkar2018generative} improve the capacity of adversarial learning by introducing advanced generative networks to provide superior reconstruction quality with extremely low bit-rates. Mentzer \textit{et al.}~\cite{mentzer2020high} demonstrated that with a hyperprior based compression model and a generative convolutional decoder with \textit{ChannelNorm}, it is possible to achieve similar visual quality on high-resolution natural images with only half the bit-rates.

In summary, the tremendous progress in learned image compression unveils the power of machine learning techniques. Nevertheless, there are still a large number of problems to investigate, which requires a systematic benchmark to illustrate critical areas where end-to-end learned frameworks for image compression can be further improved. In the following, we first analyze the important components (\textit{i.e.}, the backbone architecture and entropy model) in detail and then conduct the benchmark analysis on the methods according to various aspects.

\section{Backbones for Image Compression}
\label{sec:backbone}

A typical neural network backbone for image compression is built upon the VAE architecture. The architecture encodes images into vectors in a latent space, forming a compact representation. With dimensionality reduction and entropy constraints, the redundancy in the image is squeezed out by the compressive transform. There have been a variety of architectures for the backbone of the framework, which can be coarsely divided into two categories, namely, one-time feed-forward frameworks and multistage recurrent frameworks. Each component in a one-time feed-forward framework conducts the feed-forward operation only once in the encoding and decoding procedure. Usually, multiple models need to be trained to cover different ranges of bit-rates, as the encoder and decoder networks determine the rate-distortion trade-off. In contrast, in multistage recurrent frameworks, an encoding component of the network iteratively conducts compression on the original and residual signals, and the number of iterations controls the rate-distortion trade-off. Each iteration encodes a portion of the residual signal with a certain amount of bits. Such a model can conduct variable-bit-rate compression on its own. In the following, we introduce both types of architectures and conduct a comparison analysis on them.

\subsection{One-Time Feed-Forward Frameworks}

\begin{figure}[!t]
	\centering
	\captionsetup{labelfont={color=black},font={color=black}}
\begin{subfigure}[!h]{0.6\linewidth}
		\includegraphics[width=1\linewidth]{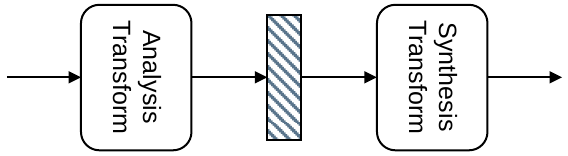}
\caption{\scriptsize GDN Transform~\cite{balle2016optimization}}
		\label{fig:cnn_sub1}
\end{subfigure}
\begin{subfigure}[!h]{0.65\linewidth}
		\includegraphics[width=1\linewidth]{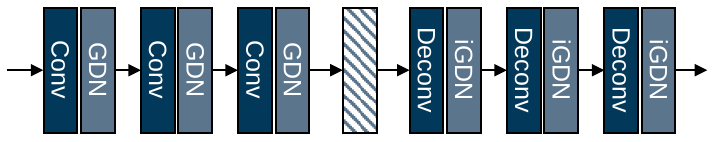}
\caption{\scriptsize Multilayer GDN Network~\cite{balle2016end}}
		\label{fig:cnn_sub2}
\end{subfigure}
\begin{subfigure}[!h]{0.6\linewidth}
		\includegraphics[width=1\linewidth]{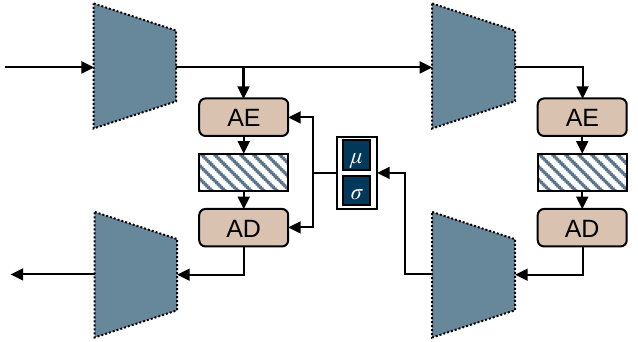}
\caption{\scriptsize Hyperprior Model~\cite{balle2018variational}\footnotemark,\cite{minnen2018joint}}
		\label{fig:cnn_sub3}
\end{subfigure}
\begin{subfigure}[!h]{0.68\linewidth}
		\includegraphics[width=1\linewidth]{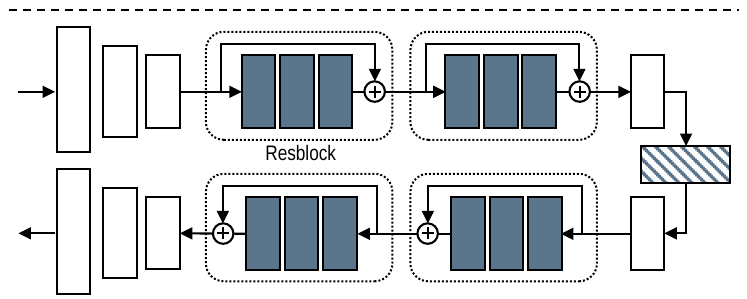}
\caption{\scriptsize Residual Auto-Encoder~\cite{theis2017lossy}}
		\label{fig:cnn_sub4}
\end{subfigure}
\begin{subfigure}[!h]{0.68\linewidth}
		\includegraphics[width=1\linewidth]{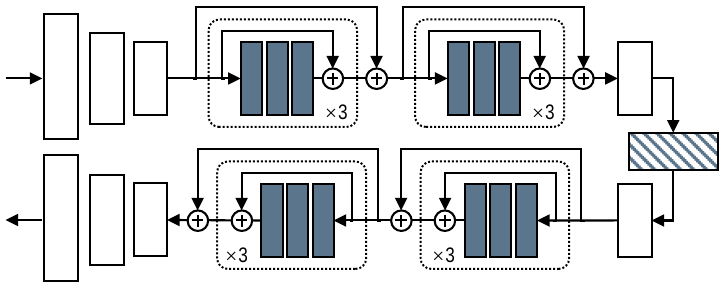}
\caption{\scriptsize Deep Residual Auto-Encoder~\cite{mentzer2018conditional}}
		\label{fig:cnn_sub5}
\end{subfigure}
\begin{subfigure}[!h]{0.68\linewidth}
		\includegraphics[width=1\linewidth]{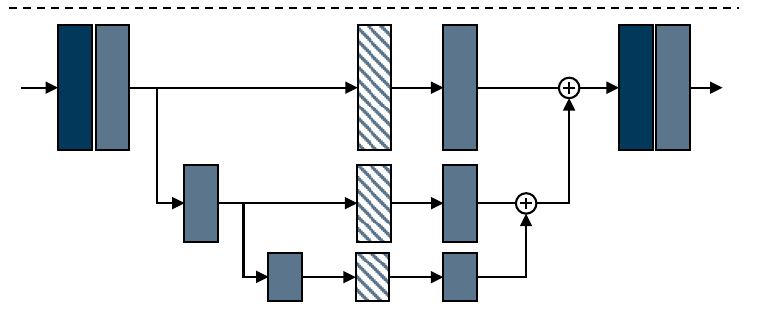}
\caption{\scriptsize Multiscale Model~\cite{nakanishi2018neural}}
		\label{fig:cnn_sub6}
\end{subfigure}
\begin{subfigure}[!h]{0.6\linewidth}
		\includegraphics[width=1\linewidth]{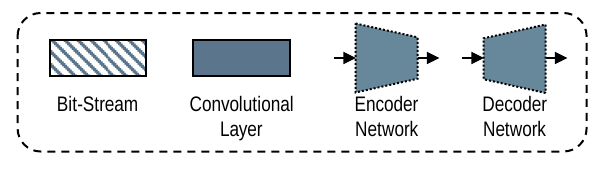}
		\label{fig:cnn_sub7}
\end{subfigure}
\caption{Illustration of typical architectures for feed-forward frameworks. The networks are divided into three categories: GDN-based networks~(a)-(c), residual block-based networks ~(d)-(e), and multiscale networks (f).}
	\label{fig:cnn}
\end{figure}

One-time feed-forward frameworks have been most widely adopted for end-to-end learned image compression. Basic variations of the architectures in the literature are illustrated in Fig.~\ref{fig:cnn}.

The first end-to-end learned image compression with a one-time feed-forward structure was proposed by Ball{\'e}~\textit{et al.}\cite{balle2016optimization}, where the analysis and synthesis transforms for encoding and decoding are made up of a single-layer GDN and inverse GDN (iGDN). This structure is then improved to support full-resolution processing, with stride convolution and the corresponding transposed convolution~\cite{balle2016end}. In later works, the hyperprior network~\cite{balle2018variational} is introduced to extract the side information from the latent representation produced by the analysis transform, and the side information can improve the entropy estimation of the latent code.

In addition to the frameworks equipped with GDN, another kind of feed-forward network utilizing residual blocks is proposed by Theis \textit{et al.}~\cite{theis2017lossy} and Mentzer \textit{et al.}~\cite{mentzer2018conditional}. These networks stack multiple residual blocks in both the encoder and decoder, greatly expanding the depth of the model. With deeper networks, the encoder and decoder can embed more complex prior images, and they have more flexibility in modeling nonlinear transforms. In addition, some works adopt a multiscale structure~\cite{rippel2017real,nakanishi2018neural}, which also extends the capacity of the network.

It is reported that a more complex design of an architecture with GDN may bring further improvements in compression performance~\cite{lee2019extended,zhou2019multi}, but not as significant as that of other contributions, such as a hyperprior.
Unlike other computer vision tasks, \textit{e.g.} image recognition, where a deeper network can usually bring extra gain in performance~\cite{he2016deep,Szegedy2015going}, it does not result in significant improvements in performance to extend the architecture complexity for learned image compression.
\footnotetext{In \cite{balle2018variational} the probability is models with zero-mean Gaussian and the prediction network only generates $\sigma$. In this case, we have $\mu \equiv 0$.}
Although deeper architectures can provide more fidelity to model the prior of the images, they are harder to train than shallower networks, especially with a hard bottleneck in the pipeline. However, with sufficient capacity, due to the characteristics of this problem, an end-to-end optimization process may easily fall into local minima, and therefore, performance is not significantly improved with increased complexity.

\subsection{Multistage Recurrent Frameworks}

\begin{figure}[!t]
	\centering
\begin{subfigure}[!h]{0.65\linewidth}
		\includegraphics[width=1\linewidth]{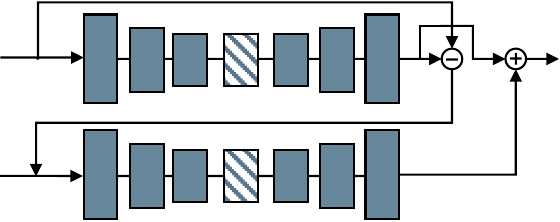}
\caption{\scriptsize Vanilla}
		\label{fig:rnn_sub1}
\end{subfigure}
\begin{subfigure}[!h]{0.65\linewidth}
		\includegraphics[width=1\linewidth]{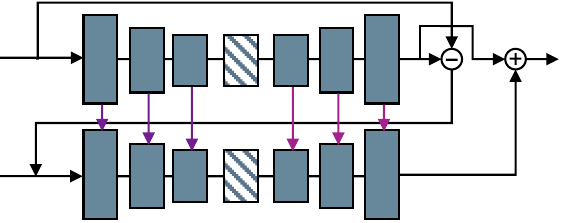}
\caption{\scriptsize Stateful}
		\label{fig:rnn_sub2}
\end{subfigure}

\begin{subfigure}[!h]{0.30\linewidth}
		\includegraphics[width=1\linewidth]{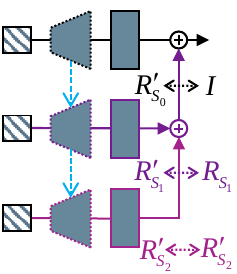}
\caption{\scriptsize Incremental}
		\label{fig:rnn_sub3}
\end{subfigure}
\begin{subfigure}[!h]{0.32\linewidth}
		\includegraphics[width=1\linewidth]{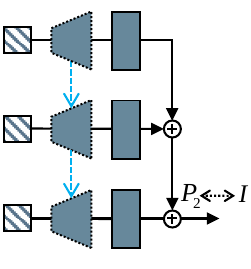}
\caption{\scriptsize Skip-Connection}
		\label{fig:rnn_sub4}
\end{subfigure}
\begin{subfigure}[!h]{0.31\linewidth}
		\includegraphics[width=1\linewidth]{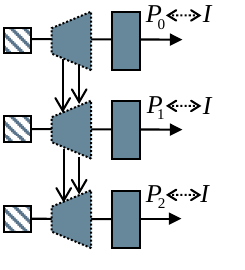}
\caption{\scriptsize Stateful Propagation}
		\label{fig:rnn_sub5}
\end{subfigure}
\begin{subfigure}[!h]{0.32\linewidth}
\end{subfigure}
\caption{Illustration of the backbones of the multistage recurrent framework and its variations. The main feature of these designs is that the residue for one stage is taken as the input at the next stage. (a) and (b) show the vanilla structure and its improved stateful form~\cite{toderici2015variable}. (c)-(e) show different cross-stage connections~\cite{baig2017learning}.}
	\label{fig:rnn}
\end{figure}

The basic architecture and the variations of multistage recurrent frameworks for image compression are illustrated in Fig.~\ref{fig:rnn}.

The vanilla multistage framework, as an illustration of the concept, progressively encodes the residue to compress the image. For an example of the simplified case, in the first stage, there is no reconstructed signal, so the residue is the original image itself. After the encoding and reconstruction, the residual image with respect to the reconstructed and original image is pushed into the network to conduct the second-stage compression. As at each stage the compression loses some of the information, the output of the second stage is the degraded signal of the \textit{true} residue. The framework compresses the residue and the residue of the residue progressively to achieve better quality. To finally reconstruct the original image, bits of all the stages are needed to decode the multistage residue maps, which are added together to form the decoded image. This kind of reconstruction process corresponds to the \textit{Incremental} structure in Fig.~\ref{fig:rnn}. The vanilla multistage framework adopts a stateless structure, where the analysis of different stages of the residue is conducted independently. It is difficult for the network to simultaneously compress the image and the residue of all steps. Therefore, in the first practical multistage structure~\cite{toderici2015variable}, a stateful framework utilizing long short-term memory (LSTM) architectures~\cite{hochreiter1997long} is introduced. LSTM maintains a state during sequential processing that propagates the features of the images to the following steps to facilitate the modeling of the multilevel residue. Fig.~\ref{fig:rnn} shows the unrolled stateful structure. In each stage, the modules in the pipeline take the currently processed residue and the state from the previous stage as the input. The states are updated and propagated for processing in the next step.

There have been studies on the aggregation of the output of each stage. Baig~\textit{et. al.}~\cite{baig2017learning} present and analyze different kinds of aggregation schemes. The basic \textit{Incremental} scheme adds the output of all stages together to form the final decoded images. The loss function of the \textit{Incremental} scheme usually includes a term to encourage the output of each stage to approximate the residue of the previous stage. A different way to combine all the stages is to treat the multistage structure as a residual network to form the \textit{Skip-Connection} scheme. There is only one term in the loss function for such a scheme to require that the sum of all the stages reconstructs the original image. Unlike the \textit{Incremental} structure, there is no explicit arrangement of the residue in the \textit{Skip-Connection} structure. The outputs of all stages contribute to the final reconstruction, each as a supplement of the reconstruction quality with respect to the others. In addition to these two kinds of schemes, Baig~\textit{et. al.} reported that with the \textit{Stateful-Propagation} structure and the corresponding residual-to-image prediction, where each step produces a prediction of the original image rather than the residual signal, the network achieves the best performance. In such a stateful propagation scheme, it is important to propagate the states of the layers to the next step to construct a refined decoding image.

\subsection{Comparative Analysis}
Each of the two categories of backbone architectures has its own properties and corresponding pros and cons. The differences are mainly due to the choice between the one-time structure and the progressive structure. Here are some main differences.
\begin{itemize}
\item Recurrent models can naturally handle variable-rate compression, while for the feed-forward network, multiple instances of networks need to be trained to support a variable range of bit-rates.
\item Feed-forward networks are comparatively shallower, and the path of back-propagation is much shorter. Training such a network can be easier. In contrast, training the recurrent models requires the back-propagation through time (BPTT) technique, which is more complicated.
\item Weights are shared across different stages in the recurrent model; thus, the total number of parameters for a practical image codec may require less storage for the parameters compared with one-time feed-forward models. However, residual signals and image signals are different in nature, making the training of a recurrent model more challenging.
\item It usually takes more time for recurrent models to encode and decode an image because the network is executed multiple times.
\end{itemize}

Despite the pros and cons, existing works report higher rate-distortion performance in one-time feed-forward architectures~\cite{johnston2018improved,minnen2018joint}. However, variable-bit-rate compression is commonly required by applications, which becomes the major barrier for end-to-end learned image compression methods to be adopted by existing systems. More efforts are needed to investigate an efficient way to achieve variable-rate compression for learning-based approaches.

\section{Entropy Models}
\label{sec:entropy}

Entropy coding is an important component in an image compression framework. According to information theory~\cite{cover2012elements}, the bit-rate needed to encode the signal is bounded by the information entropy, which corresponds to the probability distribution of the symbols in representing the signal. Thus, the entropy coding component is embedded in the end-to-end learned image compression framework to estimate the probability distribution of the latent representations and apply constraints on the entropy to reduce the bit-rate.

\begin{figure}[t]
	\centering
	\includegraphics[width=0.82\linewidth]{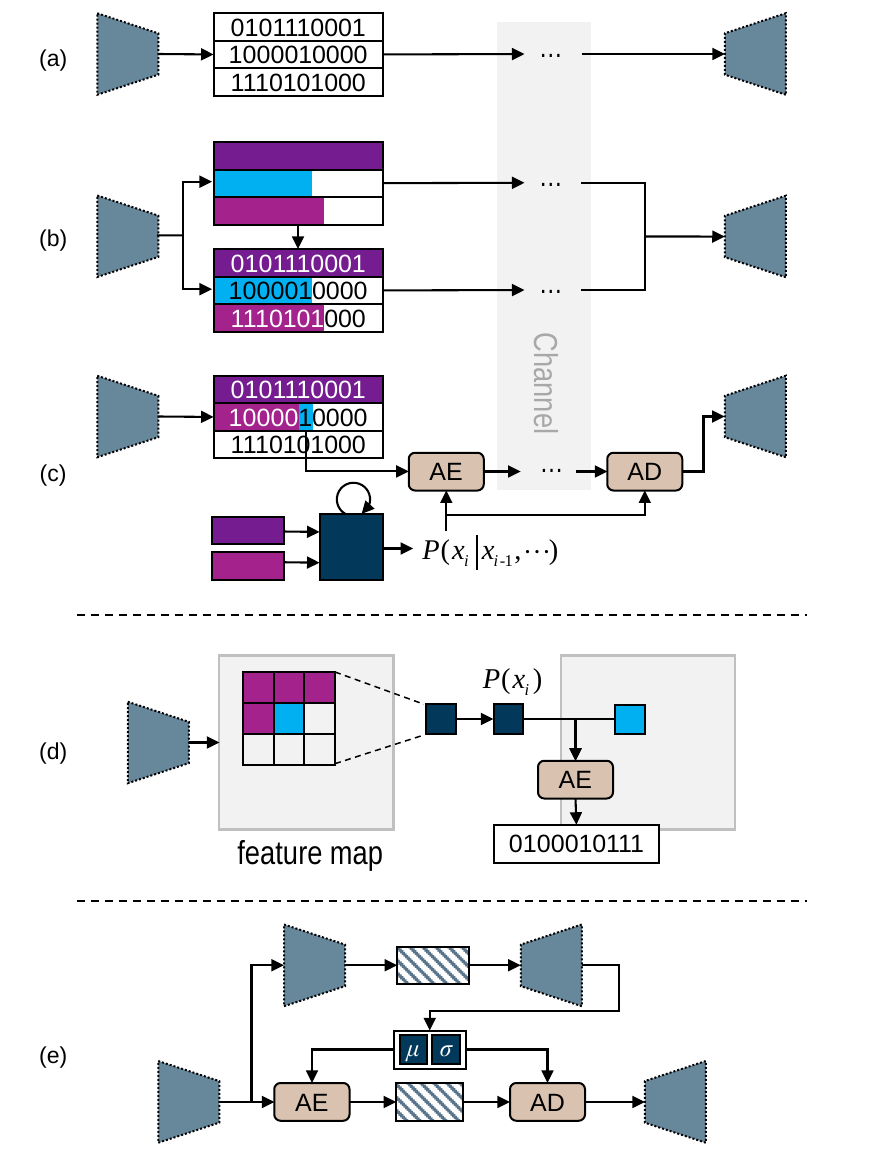}
\caption{Illustration of entropy modeling methods. (a)-(c) Binary methods and variations with the masking and context model, including (a) direct modeling~\cite{toderici2015variable}, (b) masked modeling~\cite{minnen2017spatially,li2018learning,covell2017target}, and (c) the binary context model~\cite{toderici2017full}. (d) Spatial context model for latent code maps~\cite{lee2018context,minnen2018joint,mentzer2018conditional}. (e) Hyperprior entropy model~\cite{balle2018variational}.}
	\label{fig:context}
\end{figure}

There is a large amount of research on entropy models for learned image compression. A summary of solutions to the problem of entropy modeling is presented in Table~\ref{tab:entropy}, and we illustrate the typical structure of different variations in Fig.~\ref{fig:context}.

Ideal entropy coding requires precise estimation of the joint distribution of the elements in the latent representations, for each instance of the image. In earlier works, those elements are assumed to be independently distributed~\cite{balle2016end,theis2017lossy} to simplify the design. However, even with optimized transforms, it is still difficult to eliminate the spatial redundancy in the latent maps of the images. Thus, a variety of entropy models are proposed to further reduce the redundancy in the latent code. These methods include statistical analysis over a given dataset~\cite{agustsson2017soft,theis2017lossy,balle2016end,minnen2018image,cheng2019learning}, contextual prediction or analysis~\cite{toderici2017full,covell2017target,minnen2018joint,johnston2018improved,li2018learning,nakanishi2018neural,cai2018deep,lee2018context,chen2019neural}, and utilizing a learned hyperprior~\cite{balle2018variational,lee2018context,minnen2018joint} for entropy modeling. The entropy model provides the estimation of the likelihood for all the elements, and the expectation of the log-likelihoods is the bound of the bit-rates in encoding these elements. With the entropy model, in most of the works, arithmetic coding~\cite{witten1987arithmetic} is utilized to practically losslessly encode the symbols of the latent representations.

It is worth noting that in traditional hybrid frameworks, improvements of the entropy model only affect entropy coding performance. For the learned method, as all the components are jointly optimized, a better designed entropy model not only produces a more precise estimate of the entropy but also changes the patterns produced by the analysis transform. As a consequence, the design of the entropy model should also take the structure of other components in the pipeline into consideration.

\begin{table*}[!h]
	\caption{Description of different entropy models utilized in learned image compression.}
	\label{tab:entropy}
	\centering
	\begin{tabular}{cM{7em}m{37.9em}}
		\toprule
		\multicolumn{2}{c}{Solutions} & \multicolumn{1}{M{37.9em}}{Description} \\
		\midrule
		\multirow{3}[32]{*}{Binary} & Direct & The proposed network directly produces binary codes, which are transmitted as the bit-stream without entropy modeling~\cite{toderici2015variable}. Optional external entropy codecs, such as adaptive arithmetic coding \cite{rippel2017real}, can be applied to the bit-stream to improve coding efficiency.\bigstrut   \\

		\cline{2-3}      & Masked & \bigstrut In addition to the binary code, the network also constructs a mask from the feature to indicate the length of the binary code\cite{minnen2017spatially,covell2017target,li2018learning}. The mask is usually transmitted together with the bit-stream. With rate-control, the overall performance can be further improved compared to the direct scheme.\bigstrut \\

		\cline{2-3}      & Binary Context-Model & \bigstrut The probability distribution of all the symbols to be encoded is estimated by the network with previously coded symbols~\cite{toderici2017full} and spatially adjacent symbols \cite{johnston2018improved}.  The context model can more accurately estimate the probability so that the entropy coding can be conducted with more efficiency. \\
		\midrule
		\multirow{4}[38]{*}{Statistical} & Histogram & The probability distribution is estimated by a histogram~\cite{agustsson2017soft}. A variation of this scheme is to use a Laplace-smoothed histogram for better generalization~\cite{theis2017lossy}.\bigstrut  \\
		\cline{2-3}      & Piecewise Linear  & \bigstrut  The probability density function is approximated by a parametric piecewise linear function during training~\cite{balle2016end}. Context Adaptive Binary Arithmetic Coding (CABAC)~\cite{marpe2003context} is used to practically compress the latent codes.\bigstrut  \\
		\cline{2-3}      & Parametric Factorized & \bigstrut  A function $p(x_i) = f(x_i, \theta)$ with trainable parameters $\theta$ is modeled to estimate the probability of a symbol $x_i$. These parameters reflect the distribution of latent code through the training set and can be generalized for all images~\cite{balle2018efficient,cheng2019learning}. \bigstrut   \\
		\cline{2-3}      & Gaussian (Mixture) & \bigstrut  Networks based on VAE assume that the latent code follows an elementwise Gaussian distribution. The loss function includes a term of cross-entropy between the actual distribution and the estimated Gaussian distribution to control the bit-rate~\cite{balle2018variational,minnen2018joint,lee2018context}. Gaussian Mixture distribution is shown to better estimate the likelihoods~\cite{cheng2020learned}.   \\
		\midrule
		\multirow{2}[14]{*}{Context-Model} & PixelRNN  PixelCNN & Multistage recurrent models~\cite{toderici2017full} employ PixelRNN~\cite{van2016conditional}, while one-time feed-forward models~\cite{nakanishi2018neural,mentzer2018conditional} utilize PixelCNN~\cite{oord2016pixel} for spatial context conditioned probability modeling. \bigstrut \\
		\cline{2-3}      & Masked Convolution & \bigstrut Masked 2D~\cite{lee2018context,minnen2018joint} or 3D~\cite{chen2019neural} convolutions can be seen as a simplified version of PixelCNN for conditional probability modeling. It estimates likelihoods of a to-be-encoded element based on decoded elements. \\
		\midrule
		\multirow{2}[18]{*}{Side-Information Guided} & Offline & The latent code produced by a given encoder is analyzed offline in tiles by learning a dictionary, and the indices are transmitted with lossless compression~\cite{minnen2018image}.\bigstrut  \\
		\cline{2-3}    & Hyperprior & \bigstrut The hyperprior, transmitted in the bit-stream, encodes the parameters of a Gaussian entropy model~\cite{balle2018variational} to estimate the likelihoods of the elements to be encoded. It greatly improves the accuracy of the entropy model and it can be combined with the context model for enhanced modeling. \\
		\bottomrule
	\end{tabular}%
\end{table*}

In summary,
existing methods
aim to provide a flexible transform and an accurate entropy model, all of which are neural network-based and end-to-end trainable. In addition to the main goal of rate-distortion performance, several issues need to be addressed in the exploration. The model should be adaptive to different ranges of resolutions, bit-rates, and distortions. Currently, when high-resolution capturing and displaying devices emerge, high-efficiency compression of high-resolution images is a constantly growing need. On the other hand, with the rapid development of large-scale parallel computing devices, \textit{e.g.} GPU, models should also be designed to take advantage of parallel computing devices for higher efficiency. According to the above analysis, the one-time feed-forward frameworks with convolutional neural network-powered hyperprior structures have more potential to be scalable to a high-resolution and to support large-scale parallelism. With this idea in mind, we adopt a one-time feed-forward framework and achieve
one step towards obtaining
superior performance with a newly proposed coarse-to-fine hyperprior compression model.

\section{Proposed Coarse-to-Fine Model}

\label{sec:propose}
\subsection{Coarse-to-Fine Hyperprior Modeling}

As analyzed, we follow the basic framework of a one-time feed-forward framework, which consists of an analysis transform $\mathcal{G}_a$ and a synthesis transform $\mathcal{G}_s$. $\mathcal{G}_a$ transforms the image to latent representations, and $\mathcal{G}_s$ reconstructs the image from those representations. To perform entropy coding, the latent representations are first quantized to a vector of discrete symbols $\mathbf{X} = \{X_1, X_2, ..., X_n\}$. In addition, a parametric entropy model $Q_{\mathbf{X}}(\mathbf{X}; \theta)$ \textit{w.r.t.} the random vector $\mathbf{X}$ is built to provide the estimation of the likelihoods. The aim of entropy coding is now to jointly optimize the parameters in the networks to 1) accurately model the distribution $P_{\mathbf{X}}({\mathbf{X}})$ of the random vector $\mathbf{X}$ with $Q_{\mathbf{X}}(\mathbf{X}; \theta)$ and 2) minimize the overall rate-distortion function with the estimated entropy. State-of-the-art methods combine context models and hyperpriors. In such approaches, it is first assumed that the joint probability distribution of $\mathbf{X}$ can be factorized to the product of sequential conditional probabilities as follows:
\begin{equation}
	\begin{split}
	Q_{\mathbf{X}	|\mathbf{Y}}\left(\mathbf{X}|\mathbf{Y}\right) & = \prod_i Q_{i}\left(X_i|X_{i-1},X_{i-2},...,X_{i-m}, \mathbf{Y}\right),
	\end{split}
\end{equation}
where $\mathbf{Y}$ denotes the hyperprior, which is generated from $\mathbf{X}$ and encoded to the bit-stream. When we need to decode $\mathbf{X}$, $\mathbf{Y}$ has already been decoded. These kinds of models need to address two issues. First, the dimensionality and the corresponding bit-rate of $\mathbf{Y}$ should be kept low; otherwise, $\mathbf{Y}$ itself may contain too much redundancy and is not efficiently compressed. In such a circumstance, the hyperprior may not provide enough information to accurately model the conditional probability, especially for higher ranges of bit-rates and large resolutions. Second, although contextual conditioning can help with accuracy, it is performed in a sequential way and is hard to accelerate with large-scale parallel computing devices. Thus, the framework is less scalable for input images of different sizes.

To address the issues of the sequential context models, in the proposed method, we adopt a multilayer conditioning framework, which improves scalability for images of different sizes. The formulation is modified as follows:
\begin{equation}
\label{eq:layered}
\centering
Q_{\mathbf{X}}\left(\mathbf{X}\right)=Q_{\mathbf{X},\mathbf{Y}}\left(\mathbf{X},\mathbf{Y}\right)=Q_{\mathbf{Y}}\left(\mathbf{Y}\right)Q_{\mathbf{X}|\mathbf{Y}}\left(\mathbf{X}|\mathbf{Y}\right).
\end{equation}
The first equality in Eq.~(\ref{eq:layered}) holds for $\mathbf{Y}$ because the hyperprior is generated from $\mathbf{X}$ in a deterministic manner. When $\mathbf{X}$ becomes complex and is controlled by expanding the dimension, $\mathbf{Y}$ may need to embed more information to support accurate conditional modeling. Therefore, an additional layer of the hyperprior is introduced as follows:
\begin{equation}
\label{eq:layeredz}
\centering
Q_{\mathbf{Y}}\left(\mathbf{Y}\right)=Q_{\mathbf{Y},\mathbf{Z}}\left(\mathbf{Y},\mathbf{Z}\right)=Q_{\mathbf{Z}}\left(\mathbf{Z}\right)Q_{\mathbf{Y}|\mathbf{Z}}\left(\mathbf{Y}|\mathbf{Z}\right),
\end{equation}
which in fact forms a coarse-to-fine hyperprior model. The dimension of $\mathbf{Z}$ is reduced, and the redundancy is squeezed out by the hypertransforms. Thus, the joint distribution $P_{\mathbf{Z}}(\mathbf{Z})$ of the latent representation $\mathbf{Z} = \{Z_1, Z_2,\cdot \cdot \cdot, Z_n\}$ at the innermost layer can be approximately factorized as follows:
\begin{equation}
\label{eq:z}
\centering
Q_{\mathbf{Z}}(\mathbf{Z}) = Q_{\mathbf{Z}}(Z_1, Z_2,\cdot \cdot \cdot, Z_n) \approx \prod_i Q_{Z_i}(Z_i).
\end{equation}
With Eq.~(\ref{eq:layered}) and Eq.~(\ref{eq:layeredz}), the probability distribution of $\mathbf{Y}$ and $\mathbf{X}$ can now be modeled in a conditional way, while existing works~\cite{van2016conditional,mirza2014conditional} show that neural networks are capable of modeling conditional probability distributions. The hyperrepresentation $\mathbf{Y}$ is also designed to embed the main information of the images to be compressed. Therefore, the joint distribution can also be approximately factorized as follows:
\begin{equation}
\begin{split}
Q(\mathbf{X}|\mathbf{Y}) &= Q(X_1,\cdot \cdot \cdot, X_n|\mathbf{Y}) \approx \prod_i Q_{X_i|\mathbf{Y}}(X_i|\mathbf{Y}),\\
Q(\mathbf{Y}|\mathbf{Z}) &= Q(Y_1,\cdot \cdot \cdot, Y_n|\mathbf{Z}) \approx \prod_i Q_{Y_i|\mathbf{Z}}(Y_i|\mathbf{Z}),
\end{split}
\end{equation}
where all elements in the previous layer can be utilized as the conditions to estimate the distribution of the latent representation at the upper layer. Although no contextual conditioning is conducted here, contextual conditioning can be implicitly modeled in the information flow from $\mathbf{X}$ to $\mathbf{Y}$ and then used to predict $\mathbf{X}$ from $\mathbf{Y}$. Unlike existing block-conditioning context models, in the proposed framework, the estimation of the probability for each element utilizes information from a larger area due to the coarse-to-fine structure. This helps to explore long-term correlations in images and improves the compression performance, especially for high-resolution images.

\subsection{Network Architecture}

\begin{figure*}[htbp]
	\centering
	\includegraphics[width=0.78\linewidth]{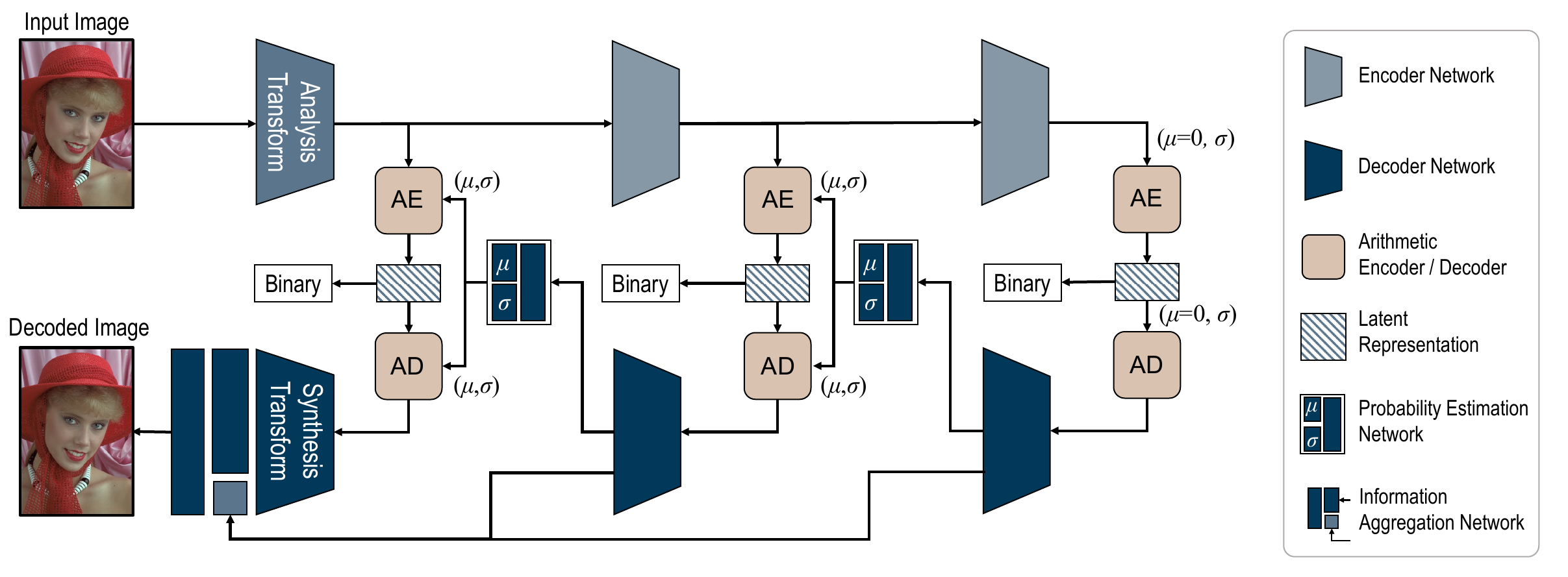}
\caption{Overall architecture of the multilayer image compression framework. The probability distribution of the innermost layer of the hyperprior is approximated with a zero-mean Gaussian distribution, where the scale values $\sigma$ are channelwise independent and spatially shared.}
	\label{fig:arch}
\end{figure*}

The overall structure of the end-to-end learned coarse-to-fine framework is shown in Fig.~\ref{fig:arch} jointly with the encoder and decoder. The analysis transform network encodes the input image as the latent representation $\mathbf{X}$, which is then quantized with a rounding operation. It aims to squeeze out pixelwise redundancy as much as possible. We exploit GDN as the activation in the analysis transform and inverse GDN in the synthesis transform. We conduct coarse-to-fine modeling with multilayer hyper analysis and a symmetric hyper synthesis transform. According to Eq.~(\ref{eq:layered}) and Eq.~(\ref{eq:layeredz}), to estimate the distribution of $\mathbf{X}$, a probability estimation network is employed to process $\mathbf{Y}$ and predict the likelihood $P_{X_i}(X_i=x_i)$ with the estimated $Q_{X_i}(X_i=x_i)$ for each element $X_i$ in $\mathbf{X}$. As stated in \cite{balle2018variational}, the conditional distribution of each element in $\mathbf{X}$ can be assumed to be Gaussian, and the probability estimation network predicts the mean and scale of the Gaussian distribution. As the latent code has been rounded to be discrete, the likelihood of the latent code can be calculated as follows:
\begin{equation}
\begin{split}
	Q_{X_i|\mathbf{Y}}(X_i&=x_i|\mathbf{Y}) = \\
	&\phi\left(\frac{x_i + \frac{1}{2} - \mu_{x_i}}{\sigma_{x_i}}\right) - \phi\left(\frac{x_i - \frac{1}{2} - \mu_{x_i}}{\sigma_{x_i}}\right),
\end{split}
\end{equation}
where $\phi$ denotes the cumulative distribution function of a standard normal distribution, while the mean $\mu_{x_i}$ and scale $\sigma_{x_i}$ are predicted from $\mathbf{Y}$. The same process is conducted \textit{w.r.t.} $\mathbf{Y}$ and $\mathbf{Z}$ to estimate the probability distribution of $\mathbf{Y}$. As illustrated in Eq.~(\ref{eq:z}), the probability distribution of $\mathbf{Z}$ can be
approximately factorized.
Thus, we employ a zero-mean Gaussian model. The likelihood of each element in $\mathbf{Z}$ can be calculated as follows:
\begin{equation}
	Q_{Z_i}(Z_i=z_i) = \phi\left(\frac{z_i + \frac{1}{2}}{\sigma_{z_i}}\right) - \phi\left(\frac{z_i - \frac{1}{2}}{\sigma_{z_i}}\right).
\end{equation}
Note that $\sigma_{z_i}$ is a trainable parameter in the network. All elements in a channel in the latent representation share the same $\sigma$ while each channel has an independent one.

According to information theory, the minimum bit-rate required to encode $\mathbf{X}$ (or $\mathbf{Y}$ and $\mathbf{Z}$) with the estimated distribution equals the cross entropy of the real distribution $P_{\mathbf{X}|\mathbf{Y}}(\mathbf{X}|\mathbf{Y})$ and the estimated distribution $Q_{\mathbf{X}|\mathbf{Y}}(\mathbf{X}|\mathbf{Y}) \sim \mathcal{N}(\mu_x, \sigma_x)$, which is denoted as follows:
\begin{equation}
\label{eq:cross}
\centering
R=H(Q)+D_{KL}(P||Q)=\mathbb{E}_{\mathbf{X}|\mathbf{Y}}\left[- \log (Q)\right].
\end{equation}
We minimize the rate-distortion function $\mathcal{L}_{RD} = R + \lambda D$ with the network. To accelerate the convergence during the training of the multilayer network, an additional information-fidelity loss is introduced. This loss term encourages the hyperrepresentation $\mathbf{Y}$ to maintain the critical information in $\mathbf{X}$
during training and is formulated as follows:
\begin{equation}
\centering
\min_{\mathbf{Y}, \theta} \mathcal{L}_{\text{if}}=||\mathcal{F}(\mathbf{Y}; \theta)- \mathbf{X}||_2.
\end{equation}
In practice, the function $\mathcal{F}$ with trainable parameter $\theta$ is one convolutional layer with no nonlinear activation. The information-fidelity loss
takes the form of the least-square error to make the prediction of $\mu$ and $\sigma$ more accurate.

\subsection{Signal-Preserving Hyper Transform}
To conduct coarse-to-fine modeling of images, especially for high-fidelity modeling in high-resolution or high-quality circumstances, it is important to preserve the information while performing hyper analysis and synthesis transforms in the succeeding hyperlayers. Therefore, the signal-preserving hypertransform is proposed to build a framework with multiple layers.
We observe that elements in the latent representations produced by the main analysis transform are much less correlated compared with pixels in natural images, as the spatial redundancy has been largely reduced by the previous analysis transforms. Therefore, local correlations in the feature maps are weak, while convolutions with large kernels rely on such local correlations for effective modeling. In addition, the previous transform network consists of stride convolutions with ReLU activation. Stride convolutions downsample the feature maps, while activation functions such as ReLU intuitively disable some of the filter neurons that produce negative values and make the response sparser. Because the dimension of these convolution layers needs to be limited to ensure the gradual factorization of the latent representation, the original hypertransform loses much information during processing.

In summary, the issues of the original analysis transform in the proposed architecture fall into two categories: 1) Original analysis transforms fix the number of channels and downsample the feature maps, which reduces the dimension of the latent maps. 2) Combining large convolution kernels with ReLUs at the beginning of the analysis transform or the end of the synthesis transform will lose some information that has not been transformed, limiting the capacity.

\begin{table}[t!]
  \centering
  \caption{Structure of the signal-preserving hypertransform.}
  \begin{subtable}[t]{1\linewidth}
  \centering
  \caption{Hyper analysis transform.}
    \begin{tabular}{cccc}
    \toprule
    Name  & Operation & Output Shape & Activation \\
    \midrule
    Input & /     & $(b, h, w, c)$ & / \\
    
    \#1 E & Conv. $(3\times3)$ & $(b, h, w, 2c)$ & Linear \\
    
    Down  & Space-to-Depth & $(b, h/2, w/2, 8c)$ & / \\
    
    \#2 E & Conv. $(1\times1)$& $(b, h/2, w/2, 4c)$ & ReLU \\
    
    \#3 E & Conv. $(1\times1)$& $(b, h/2, w/2, 4c)$ & ReLU \\
    
    \#4 E & Conv. $(1\times1)$& $(b, h/2, w/2, c)$ & Linear \\
    \bottomrule
    \end{tabular}%
  \end{subtable}
  \begin{subtable}[t]{1\linewidth}
  \centering
  \vspace{2mm}
  \caption{Hyper synthesis transform.}
    \begin{tabular}{cccc}
    \toprule
    Name  & Operation & Output Shape & Activation \\
    \midrule
    Input & /     & $(b, h/2, w/2, c)$ & / \\
    
    \#1 D & Deconv. $(1\times1)$ & $(b, h/2, w/2, 4c)$ & Linear \\
    
    \#2 D & Deconv. $(1\times1)$ & $(b, h/2, w/2, 4c)$ & ReLU \\
    
    \#3 D & Deconv. $(1\times1)$ & $(b, h/2, w/2, 4c)$ & ReLU \\
    Up  & Depth-to-Space & $(b, h, w, c)$ & / \\
    \#4 D & Deconv. $(3\times3)$ & $(b, h, w, c)$ & Linear \\
    \bottomrule
    \end{tabular}%
  \end{subtable}
  \label{tab:trans}%
\end{table}%

The signal-preserving hypertransform is designed to facilitate the multilayer structure by preserving information for coarse-to-fine analysis. The structure of the analysis and synthesis transform network is illustrated in Table~\ref{tab:trans}. Instead of using large kernels in the filters, we employ a relatively small filter in the first layer with no nonlinear activation, and we conduct $1\times 1$ convolutions in the remaining layers. The first layer in the network expands the dimension of the original representations. Combined with succeeding nonlinear layers, the expansion of dimension preserves the information of the original representations while supporting nonlinear modeling. We exploit a \textit{space-to-depth} operation to reshape the tensor of the representations, making spatially adjacent elements scatter in one location but in different channels. In this way, the succeeding $1\times 1$ convolutions are able to conduct nonlinear transforms to reduce spatial redundancy. At the final layer of the network, we conduct a dimensionality reduction on the tensor to make the representation compact. We symmetrically design the hyper synthesis transform to produce $\mathbf{Y}$ in Eq.~(\ref{eq:layered}) as the conditional prior for the outer layer, which is taken as the side information for reconstruction.

\subsection{Information Aggregation for Reconstruction}

\begin{figure}[t]
    \centering
    \includegraphics[width=0.95\columnwidth]{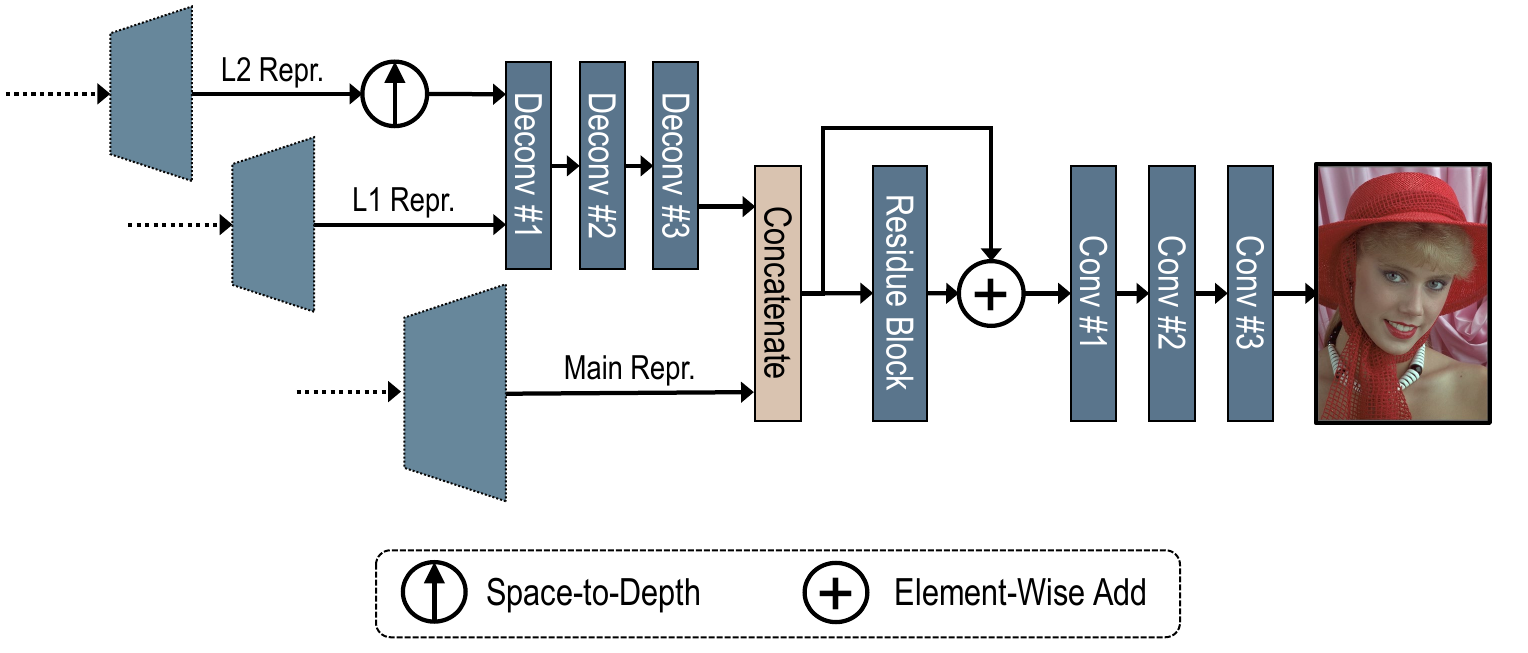}
\caption{Information aggregation subnetwork for the reconstruction of the decoded image. The main latent representation (Main Repr.) and the two layers of hyperrepresentations (L1 Repr. and L2 Repr.) are aggregated for the reconstruction.}
    \label{fig:agg}
\end{figure}

In the decoding process, the synthesis transform maps latent representations back to pixels. To best reconstruct the image, the decoder needs to fully utilize the provided information in the bit stream. Practical image and video compression usually exploit side information to improve quality. With this idea in mind, we take hyperlatent representations as side information and aggregate information from all layers of the hyperlatent representations to reconstruct the decoded image in the proposed framework. The architecture of the information aggregation decoding network is shown in Fig.~\ref{fig:agg}. Both the main latent representation and the higher order representations of smaller scales are upsampled by the decoding network to half the size of the output image. A fusion is conducted with a concatenation of the two representations. The fused representation is then processed by a residue block and then upsampled to the scale of the output image.

By fusing the main representation and the hyperrepresentations, information of different scales contributes to the reconstruction of the decoded image, where the higher-order representations provide global information and the others preserve details in the image. The fusion process is conducted at smaller spatial resolutions to avoid high computational complexity. After the fusion of features, we employ a single residue block with peripheral convolution layers to map the feature maps back to pixels.

\subsection{Implementation Details}
We follow the same network structure and hyperparameters as~\cite{balle2018variational} in the analysis and synthesis transforms, while the last layer of the original synthesis transform is removed. For the probability estimation network and the information-aggregation reconstruction network, we summarize the key properties in Table~\ref{tab:prediction} and Table~\ref{tab:iar}, respectively. We train the network with the rate-distortion tradeoff, specified in Eq.~(\ref{eq:rdo}). We train multiple models with different $\lambda$ to encode images to different bit-rates. Models optimized with MSE loss function are trained with $\lambda \in \{1.2\times 10^{-3}, 1.5\times 10^{-3}, 2.5\times 10^{-3}, 8\times 10^{-3}, 1.5\times 10^{-2}, 2.0\times 10^{-2}, 3.0\times 10^{-2}\}$, and those optimized with MS-SSIM loss function are trained with $\lambda \in \{10, 25, 45, 70, 100, 200, 300, 360\}$. The typical width (number of channels) of the transforms is set to $c = 128$, corresponding to Table~\ref{tab:trans}, and $c = 192$ for the main analysis and synthesis transforms. To provide enough degrees of freedom for high bit-rates compression, we double the width of the network when $\lambda \geq 8\times 10^{-3}$ for MSE or $\lambda \geq 70$ for MS-SSIM. 

\begin{table}[t]
	\centering
	\captionsetup{labelfont={color=black},font={color=black}}
	\caption{Structure of the probability estimation network. K and S are short for kernel size and stride, respectively. After the split, one half of the tensor is used as \textit{mean} in the Gaussian distribution. We calculate the absolute value of the other half as \textit{scale}.}
	\label{tab:prediction}
	  \begin{tabular}{cccccc}
	  \toprule
	  Layer & In Shape & Out Shape & K     & S     & Activation \\
	  \midrule
	  Unfold & $c,h,w$ & $c,5,5,h,w$ & /   & /   & None \\
	  Transpose & $c,5,5,h,w$ & $h,w,c,5,5$ & /   & /   & None \\
	  Reshape & $h,w,c,5,5$ & $h\cdot w,c,5,5$ & /   & /   & None \\
	  Conv  & $h\cdot w,c,5,5$ & $h\cdot w,c,5,5$ & 3     & 1     & Leaky ReLU \\
	  Conv  & $h\cdot w,c,5,5$ & $h\cdot w,c,3,3$ & 3     & 2     & Leaky ReLU \\
	  Conv  & $h\cdot w,c,3,3$ & $h\cdot w,c,3,3$ & 3     & 1     & Leaky ReLU \\
	  Reshape & $h\cdot w,c,3,3$ & $h\cdot w,c\cdot 9$ & /   & /   & None \\
	  Dense & $h\cdot w,c\cdot 9$ & $h\cdot w,c\cdot 2$ & /   & /   & None \\
	  Reshape & $h\cdot w,c\cdot 2$ & $h,w,c\cdot 2$ & /   & /   & None \\
	  Split & $h,w,c\cdot 2$ & $h,w,c,2$ & /   & /   & None \\
	  \bottomrule
	  \end{tabular}%
\end{table}

\begin{table}[t]
	\centering
	\captionsetup{labelfont={color=black},font={color=black}}
	\caption{Structure of the Information-Aggregation Reconstruction network, corresponding to Fig.~\ref{fig:agg}. R \#1 refers to the first layer in the residual block. C-In and C-Out refer to the numbers of input and output channels, respectively. K and S are short for kernel size and stride, respectively.}
	\label{tab:iar}
	\begin{tabular}{ccccccc}
		\toprule
		\#Layer & Layer Type & C-In & C-Out & K & S & Activation \\
		\midrule
		\#1   & Deconv. & 256   & 192   & 5   & 2     & None \\
		\#2   & Deconv. & 192   & 192   & 5   & 2     & Leaky ReLU \\
		\#3   & Deconv. & 192   & 192   & 5   & 2     & Leaky ReLU \\
		\#4   & Deconv. & 384   & 64    & 5   & 2     & None \\
		\#5   & Conv. & 64    & 3     & 3   & 1     & Leaky ReLU \\
		\#6   & Conv. & 3     & 3     & 1   & 1     & None \\
		R \#1 & Conv. & 384   & 192   & 3   & 1     & Leaky ReLU \\
		R \#2 & Conv. & 192   & 192   & 3   & 1     & Leaky ReLU \\
		R \#3 & Conv. & 192   & 384   & 3   & 1     & Leaky ReLU \\
		\bottomrule
	\end{tabular}
\end{table}

The network is trained on DIV2K~\cite{Agustsson_2017_CVPR_Workshops} dataset. The dataset contains 800 lossless images with 2K resolution on average. We down-sample the original images to half of their resolutions as an augmentation. In each training iteration, we randomly sample $256\times 256$ patches from images. We adopt a multi-step training strategy. We first pre-train the main analysis and synthesis transforms for 200,000 iterations. After that, we fix the parameters of the main transforms and train the fine-grained hyper transforms progressively. Each group of hyper transforms (\textit{i.e.} the fine-grained groups and the coarse-grained groups of hyper transforms) is trained for 20,000 iterations. Next, we train the Information-Aggregation Reconstruction subnetwork for another 20,000 iterations. Finally, we end-to-end tune the whole network for 400,000 iterations to complete the training.

\section{Evaluation}
\label{sec:evaluation}
\subsection{Datasets}

End-to-end learned image compression is a self-supervised problem where distortion metrics measure the difference between the original image and the reconstructed image and the bit-rate corresponds to the entropy of the latent code. Thus, no extra labeling labor is needed, and many existing large-scale image sets, \textit{e.g.} ImageNet~\cite{russakovsky2015imagenet} and DIV2K~\cite{Agustsson_2017_CVPR_Workshops}, can be used to train networks for image compression. To reduce possible compression artifacts in the images, the lossy-compressed images are usually downsampled before they are used for network training.

Commonly used testing image sets include Kodak~\cite{kodak} and Tecnick~\cite{asuni2014testimages}, which contain high-quality natural images that have not been lossy-compressed. The Kodak dataset consists of 24 images with resolution $512\times 768$, with a wide variety of content and textures that are sensitive to artifacts. Thus, it has been widely used to evaluate image compression methods. For the Tecnick dataset, the \textit{SAMPLING} testset is used for evaluation in some works. In contrast to Kodak, this dataset contains images with higher resolution ($1200 \times 1200$), which can serve as a supplemental benchmark for image compression methods that can have different performance on images with different resolutions.

In addition, in recent years, the CVPR Workshop and Challenge on Learned Image Compression (CLIC), with the goal of encouraging research in learning-based image compression, has attracted much attention in the community. A testing dataset consisting of images captured by both mobile phones and professional cameras is provided and updated year by year. The images have higher resolutions, on average $1913\times1361$ for \textit{mobile} photos and $1803\times1175$ for \textit{professional} photos. Evaluation results on this dataset indicate compression performance on images with relatively high resolutions.

\subsection{Rate-Distortion Performance}
\label{sec:rdperformance}
Although the overall history of the development of end-to-end learned image-compression methods is not as long as that of hybrid coding standards, there have been a significant number of works on this topic, and tremendous progress has been made. However, few studies have thoroughly evaluated rate-distortion performance on various images and compared baselines (\textit{i.e.}, anchors). It is nevertheless valuable to compare performance on technical merits to investigate which direction truly affects performance. In the following, we summarize the performance of selected works. The contributions in these works include different methods for entropy modeling, novel architecture design and normalization.

\subsubsection{Evaluation Protocol}
\begin{figure*}[!h]
	\centering
	\captionsetup{labelfont={color=black},font={color=black}}
\begin{subfigure}[t]{0.45\textwidth}
		\includegraphics[width=1\linewidth]{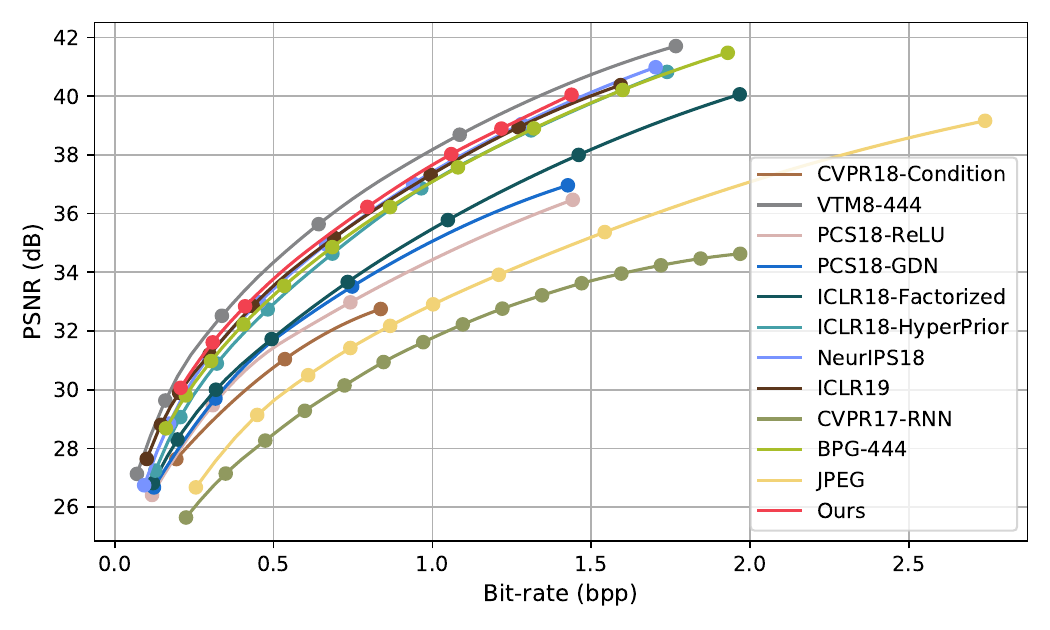}
\caption{Kodak, PSNR}
		\label{subfig:1}
\end{subfigure}
\begin{subfigure}[t]{0.45\textwidth}
		\includegraphics[width=1\linewidth]{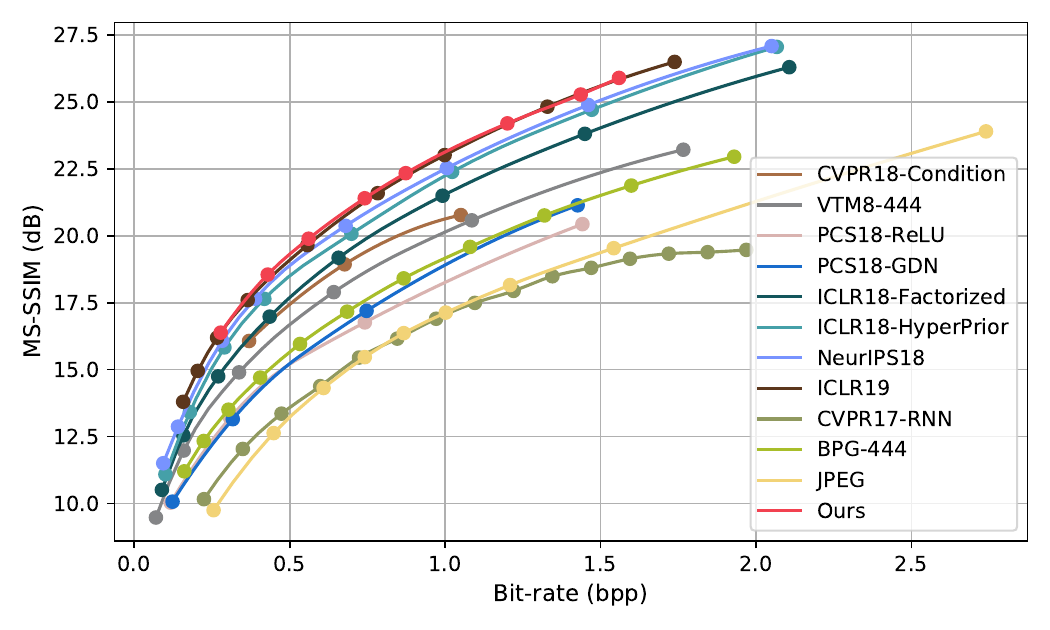}
\caption{Kodak, MS-SSIM}
		\label{subfig:2}
\end{subfigure}

\begin{subfigure}[t]{0.45\textwidth}
		\includegraphics[width=1\linewidth]{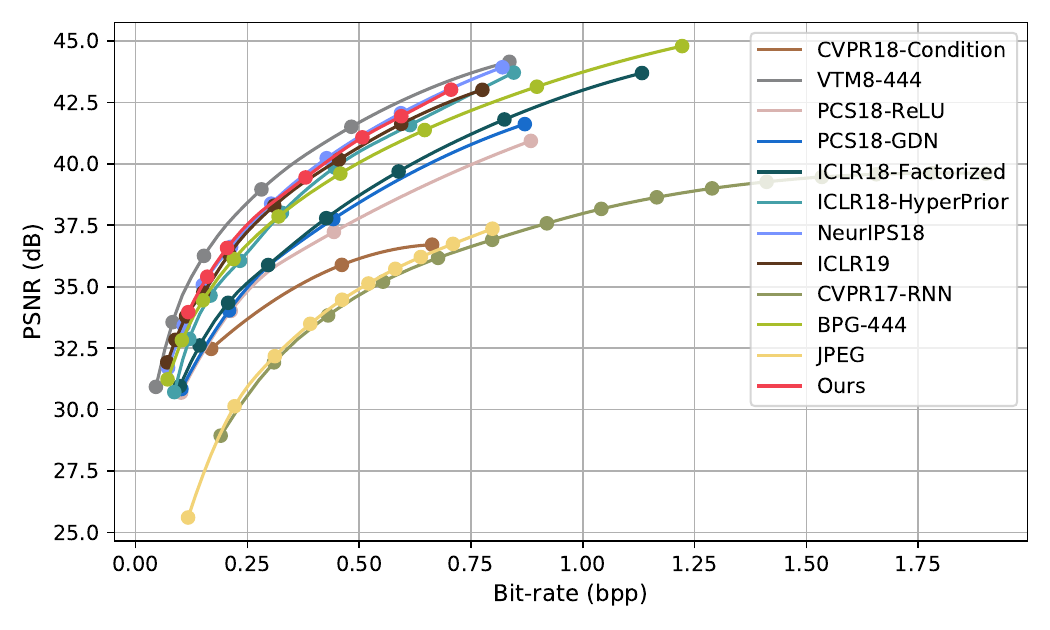}
\caption{Tecnick, PSNR}
		\label{subfig:3}
\end{subfigure}
\begin{subfigure}[t]{0.45\textwidth}
		\includegraphics[width=1\linewidth]{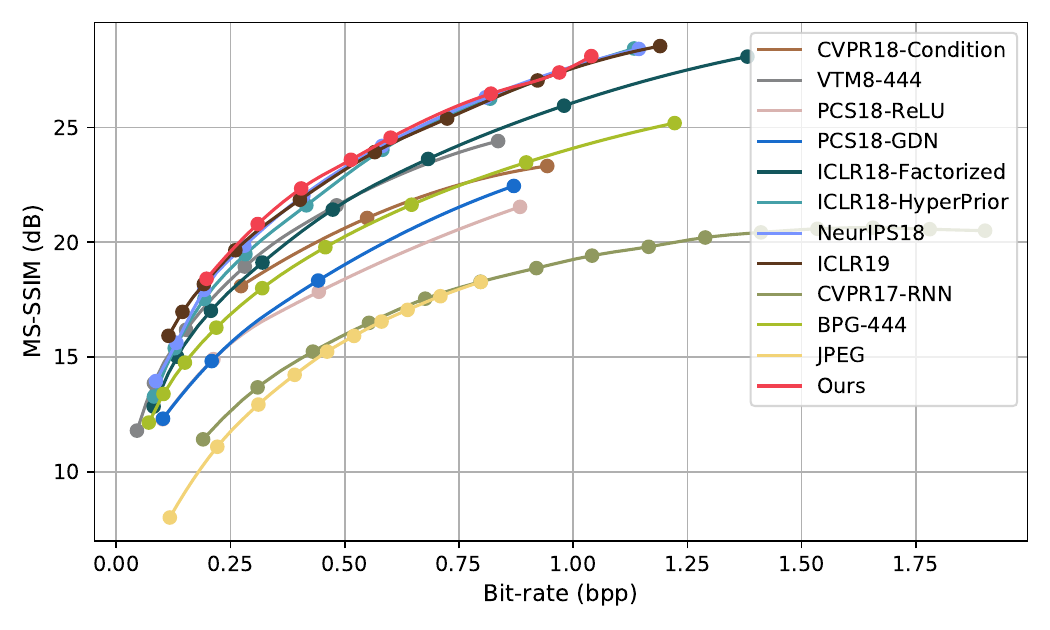}
\caption{Tecnick, MS-SSIM}
		\label{subfig:4}
\end{subfigure}

\begin{subfigure}[t]{0.45\textwidth}
		\includegraphics[width=1\linewidth]{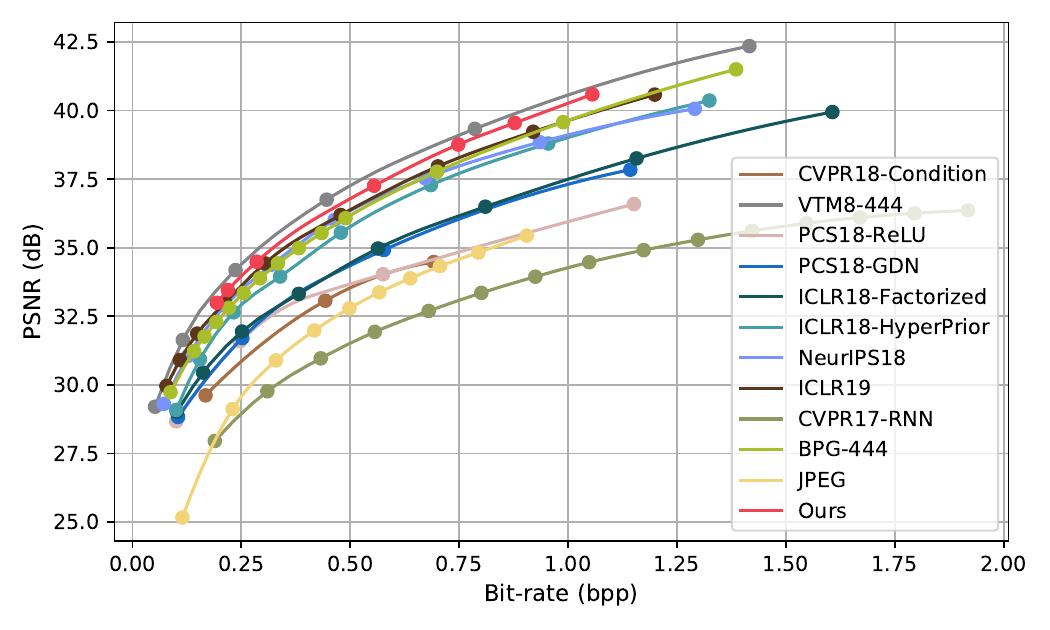}
\caption{CLIC, PSNR}
		\label{subfig:5}
\end{subfigure}
\begin{subfigure}[t]{0.45\textwidth}
		\includegraphics[width=1\linewidth]{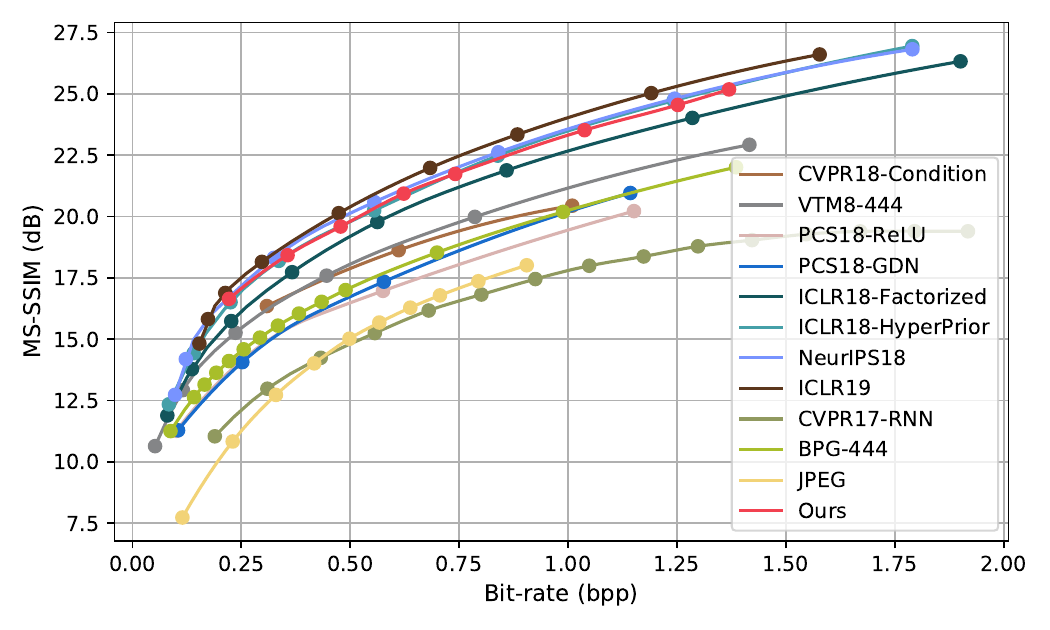}
\caption{CLIC, MS-SSIM}
		\label{subfig:6}
\end{subfigure}
\caption{
Rate-Distortion Curves. The methods include PCS18-ReLU, PCS18-GDN~\cite{balle2018efficient}, ICLR18-Factorized, ICLR18-HyperPrior~\cite{balle2018variational}, NeurIPS18~\cite{minnen2018joint}, ICLR19~\cite{lee2018context}, CVPR17-RNN~\cite{toderici2017full}, CVPR18-Condition~\cite{mentzer2018conditional}, BPG-4:4:4~\cite{bpg}, VTM8-4:4:4~\cite{vtm8} and JPEG~\cite{marcellin2000overview}. We conduct the evaluation on the three datasets Kodak, Tecnick and CLIC 2019 (validation set). PSNR and MS-SSIM are used as the distortion metrics. We convert the MS-SSIM values to decibels ($-10\log_{10}(1-d)$, where $d$ refers to the MS-SSIM value) for a clear illustration, following \cite{balle2018variational}.
}
	\label{fig:compare}
\end{figure*}

Three datasets, \textit{i.e.}, Kodak, Tecnick and CLIC 2019 validation set, are used in the evaluation corresponding to three different levels of resolution and different content. For the evaluated learning-based methods, we average the metrics of the bit-rate (bpp) and the distortion (PSNR and MS-SSIM) across the dataset for different models, which are usually trained with different trade-off coefficients $\lambda$. We compare the learning-based methods with JPEG, BPG, and VVC. For these hybrid codecs, the metrics are averaged at different quality factors (QFs) or quantization parameters (QPs). To illustrate the comparison, we show the results for rate-distortion curves in Fig.~\ref{fig:compare} \footnotemark[2]. We also calculate the BD-rate~\cite{bjontegarrd2001calculation} with respect to the bit-rate and PSNR over the three datasets. As not all methods cover the whole range of bit-rate and PSNR, we separate different bit-rate ranges for evaluation and comparison, marked as low, median, high, and full. Bit-rate ranges are different among datasets due to variations of content, but full ranges are selected to cover the variation of image quality from poor to transparent, as shown in Table~\ref{tab:bdinfo}. The BD-Rate results are shown in Table~\ref{tab:bdrate}.
We analyze the results and summarize the important properties in the following.

\begin{table}[htbp]
	\centering
    \captionsetup{labelfont={color=black},font={color=black}}
    \caption{Specifications of BD-Rate range on different testing datasets. Full ranges are illustrated, and low, median, high ranges are chosen respectively within the full ranges.}
    \label{tab:bdinfo}
	\begin{tabular}{ccc}
		\toprule
		Dataset &Bit-Rate Range &PSNR Range \\
		\midrule
		Kodak &0.25 bpp - 1.40 bpp &26 dB - 40 dB \\
		Tecnick &0.12 bpp - 0.70 bpp &26 dB - 43 dB \\
		CLIC &0.20 bpp - 1.05 bpp &28 dB - 40 dB \\
		\bottomrule
	\end{tabular}
\end{table}

\begin{table}[]
    \centering
    \captionsetup{labelfont={color=black},font={color=black}}
    \caption{Evaluation of the BD-Rate on different methods (optimized by PSNR) on different image sets. We set BPG-4:4:4 as the anchor. The negative values reflect the average bit-rate saving compared to the anchor at the same level of distortion. Best performances are marked in \textbf{bold}, while the second best ones are \underline{underlined}.}
    \label{tab:bdrate}
	\begin{tabular}{cccccc}
		\toprule
		\multirow{2}[4]{*}{Method} &\multicolumn{4}{c}{Bit-Rate Range} \\\cmidrule{2-5}
		&  Low & Median & High & Full \\
		\midrule
		\multicolumn{5}{c}{Kodak} \\
		\cmidrule{1-5}
		CVPR18-Condition &76.18\% &N/A &N/A &N/A \\
		VTM8-4:4:4 &\textbf{-20.85\%} &\textbf{-18.26\%} &\textbf{-15.86}\% &\textbf{-18.91\%} \\
		PCS18-GDN &42.09\% &40.64\% &41.62\% &41.68\% \\
		ICLR18-Factorized &35.46\% &32.75\% &28.59\% &32.96\% \\
		ICLR18-HyperPrior &6.11\% &3.06\% &0.32\% &4.07\% \\
		NeurIPS18 &-4.63\% &-4.62\% &-5.03\% &-4.68\% \\
		ICLR19 &-6.27\% &-4.75\% &-4.18\% &-5.40\% \\
		CVPR17-RNN &189.38\% &174.02\% &193.28\% &176.16\% \\
		JPEG &N/A &113.28\% &104.99\% &N/A \\
		Ours &\ul{-10.65\%} &\ul{-8.74\%} &\ul{-8.31\%} &\ul{-9.42\%} \\
		\midrule
		\multicolumn{5}{c}{Tecnick} \\
		\cmidrule{1-5}
		CVPR18-Condition &N/A &123.13\% &N/A &N/A \\
		VTM8-4:4:4 &\textbf{-31.63\%} & \textbf{-30.14\%} & \textbf{27.56\%} & \textbf{-30.25\%}\\
		PCS18-GDN &49.28\% &43.57\% &32.34\% &43.00\% \\
		ICLR18-Factorized &41.29\% &37.64\% &25.58\% &36.70\% \\
		ICLR18-HyperPrior &5.64\% &-0.23\% &-7.82\% &0.98\% \\
		NeurIPS18 &-12.76\% &\ul{-14.84\%} &\ul{-18.16\%} &\ul{-14.54\%} \\
		ICLR19 &-9.85\% &-10.92\% &-11.78\% &-11.65\% \\
		CVPR17-RNN &N/A &210.12\% &224.64\% &N/A \\
		JPEG &222.30\% &193.80\% &187.90\% &198.24\% \\
		Ours &\ul{-14.82\%} &-14.56\% &-16.48\% &-14.15\% \\
		\midrule
		\multicolumn{5}{c}{CLIC\footnotemark[3]}\\
		\cmidrule{1-5}
		CVPR18-Condition &88.73\% &N/A &N/A &N/A \\
		VTM8-4:4:4 &\textbf{-23.43\%} &\textbf{-20.31\%} &\textbf{-17.52\%} &\textbf{-21.22\%} \\
		PCS18-GDN &53.34\% &53.53\% &54.26\% &53.54\% \\
		ICLR18-Factorized &49.37\% &49.45\% &50.85\% &49.64\% \\
		ICLR18-HyperPrior &12.00\% &8.97\% &9.99\% &10.63\% \\
		NeurIPS18 &-3.53\% &-1.40\% &4.45\% &-1.88\% \\
		ICLR19 &-7.52\% &-4.33\% &-2.17\% &-4.61\% \\
		JPEG &124.30\% &115.73\% &N/A &N/A \\
		Ours &\ul{-14.49\%} &\ul{-12.21\%} &\ul{-11.64\%} &\ul{-12.86\%} \\
		\bottomrule
		\end{tabular}

\end{table}

\footnotetext[2]{The results of NeurIPS18 correspond to the publicly released code, which does not include auto-regressive context model.}

\subsubsection{Entropy Model}
The design of the entropy model is the main driving force of improvements in rate-distortion performance. The design of entropy models in end-to-end learned image compression has developed through the period from contextual binary entropy models~\cite{toderici2017full} to hyperprior models and spatial~/~cross-channel entropy estimation~\cite{balle2018variational,minnen2018joint}. Specifically, as shown in Fig.~\ref{fig:compare}, a leap in gain occurred with the emergence of hyperpriors, which have been adopted by many other frameworks. Despite great success, modeling contextual probability is still a challenging topic in image modeling due to variation in resolution. As shown in Table~\ref{tab:bdrate}, the context model-based method~\cite{lee2018context} may have unstable gain over BPG at different levels of resolution, while the proposed methods achieve consistent superiority over the anchor.

\subsubsection{Depth of the Network}
The depth of the network is a comparatively less important factor in performance, while in other computer vision tasks, networks with a deeper architecture usually perform better than those with fewer layers. Some works~\cite{zhou2019end,lee2019extended} also confirm this observation. Instead of building complicated network architectures, work may focus more on the specific design of the networks to better model the image prior. However, it has been reported that the network should consist of a sufficient quantity of parameters, and the width of the network should be large enough for effective modeling of images, especially for higher ranges of bit-rates and higher quality~\cite{lee2018context}.

\subsubsection{Normalization}
It is reported in \cite{balle2018variational} that batch normalization~\cite{Sergey2015batch}, commonly used to improve the performance of neural networks, does not bring significant improvement. However, Ball{\`e} \textit{et al.} proposed generalized divisive normalization~\cite{balle2015density,balle2018efficient}, which is proven to be able to decorrelate the elements in images to improve overall performance. Most state-of-the-art solutions adopt normalization and its inverse in the main encoding and decoding transform. However, it still remains as a topic in future research to reduce spatial redundancy more efficiently with normalization.

\subsubsection{Summary}
We evaluate the rate-distortion performance of different methods developed in recent years. As we can see from the results, great progress has been made to improve the rate-distortion performance, where the decorrelation normalization and the hyperprior model bring significant improvement. Nevertheless, we also see large variations in performance on different testing datasets. Compared with existing works, the proposed method achieves a more consistent gain on different content and resolutions.

\footnotetext[3]{The results of CVPR17-RNN on CLIC 2019 validation dataset is not included, as the available code does not support the resolutions in this dataset.}

\subsection{Studies on the Proposed Method}
\subsubsection{Coarse-to-Fine Modeling}
We propose the coarse-to-fine hyperprior model to reduce the bit-rate. We conduct ablation studies to evaluate the coarse-to-fine design. In this experiment, we benchmark on a subset from the LIU4K dataset\cite{Liu4K}, to evaluate the performance at different resolutions but having the same content. Images in LIU4K dataset are of 4K resolutions. We down-sample the images to 1080p ($1920 \times 1080$) and 540p ($960 \times 540$) resolutions to build three subsets of different resolutions. We calculate BD-Rate on the R-D curves, with the single-layer hyperprior model as the anchor. The BD-Rates results are shown in Table~\ref{tab:ablation-coarse}. As shown, the coarse-to-fine model achieve R-D performance improvements over the original single layer model. We also show that the coarse-to-fine models achieves more significant improvements in BD-Rate reduction on high-resolution images. It especially benefits emerging high-resolution applications.

\begin{table}[t]
	\centering
    \captionsetup{labelfont={color=black},font={color=black}}
    \caption{BD-Rate evaluation for the coarse-to-fine hyperprior model at different resolutions, with the single-layer hyperprior as the anchor.}
    \label{tab:ablation-coarse}
	\begin{tabular}{cc}
		\toprule
		Resolution &BD-Rate\\
		\midrule
		4K & -4.65\%\\
		1080p & -2.65\%\\
		540p & -1.97\%\\
		\bottomrule
	\end{tabular}
\end{table}

\subsubsection{Information-Aggregation Reconstruction}
\begin{figure}[t]
	\centering
	\captionsetup{labelfont={color=black},font={color=black}}
	\includegraphics[width=1\linewidth]{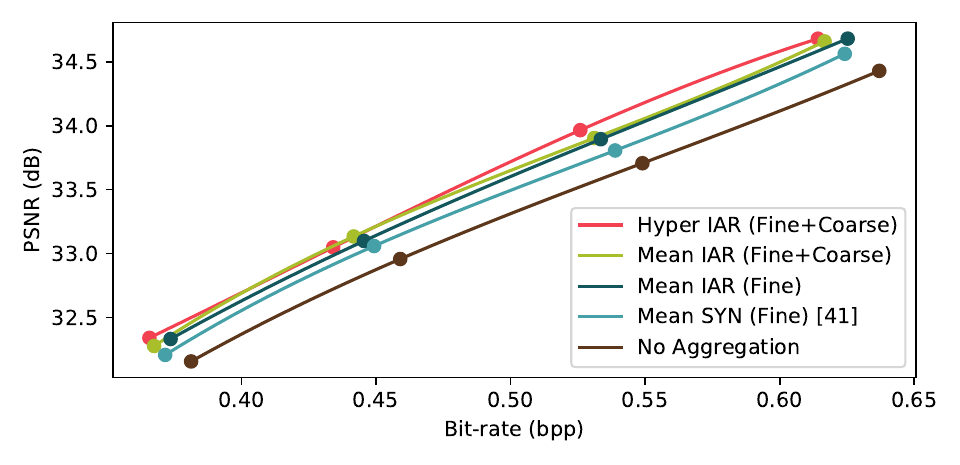}
	\caption{Rate-Distortion curves by different aggregation methods. The methods vary in aggregated feature forms (\textit{Hyper} and \textit{Mean}), feature granularity (\textit{Fine} and \textit{Fine+Coarse}), and fusion stage (\textit{SYN} and \textit{IAR}).}
	\label{fig:iar-ablation}
\end{figure}

The Information Aggregation Reconstruction (IAR) subnetwork is designed to improve reconstruction quality. It aggregates image representations at different granularities to fully utilize transmitted information for reconstruction. To analyze the effect of the IAR component, we conduct ablation studies considering the forms and granularities of the aggregated features. The results are shown in Fig~\ref{fig:iar-ablation} and Table~\ref{tab:iar-ablation}. There are two types of feature forms, \textit{i.e.} \textit{Hyper} information retrieved right after the hyper synthesis transforms, and \textit{Mean} of Gaussian distributions generated by the prediction subnetwork~\cite{zhou2019multi}. These feature maps can be aggregated at different resolutions, \textit{i.e.} at the small-resolution stage before the synthesis transform (\textit{SYN}), or at the full-resolution stage within the \textit{IAR} subnetwork. With the proposed coarse-to-fine hyperprior model, the hyperpriors can be aggregated at different granularities, \textit{i.e.} \textit{Fine} and \textit{Coarse}. We empirically analyze the effect of combining these factors in the ablation study. As shown in Fig.~\ref{fig:iar-ablation}, the fusion of multi-resolution representation shows significant benefits, and it is beneficial to aggregate information at both coarse and fine granularities. Utilizing hyperprior representation tends to show better performance than concatenating \textit{Mean} information. Besides, an aggregation at the stage of higher resolution leads to improved performance.

\begin{table}[htbp]
	\centering
	\captionsetup{labelfont={color=black},font={color=black}}
	\caption{BD-Rate corresponding with R-D curves in Fig.~\ref{fig:iar-ablation}. We use setting “No Aggregation” as the anchor.}
	\label{tab:iar-ablation}
	\begin{tabular}{ccc}
		\toprule
		Settings &BD-Rate \\
		\midrule
		Hyper IAR (Fine+Coarse) &-8.38\% \\
		Mean IAR (Fine+Coarse) &-7.34\% \\
		Mean IAR (Fine) &-6.35\% \\
		Mean SYN (Fine)~\cite{zhou2019multi} &-4.17\% \\
		\bottomrule
	\end{tabular}
\end{table}

\subsubsection{Visual Quality Analysis}
\begin{figure*}[htbp]
\centering
\captionsetup{labelfont={color=black},font={color=black}}
    \begin{subfigure}[h]{0.21\linewidth}
    \centering
      \includegraphics[width=0.9\linewidth]{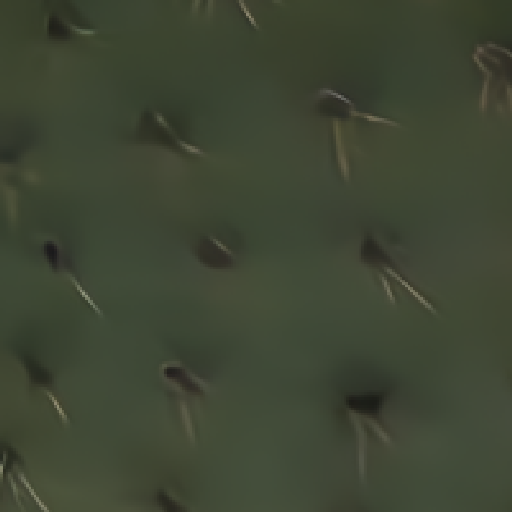}
    \end{subfigure}
    \begin{subfigure}[h]{0.21\linewidth}
    \centering
      \includegraphics[width=0.9\linewidth]{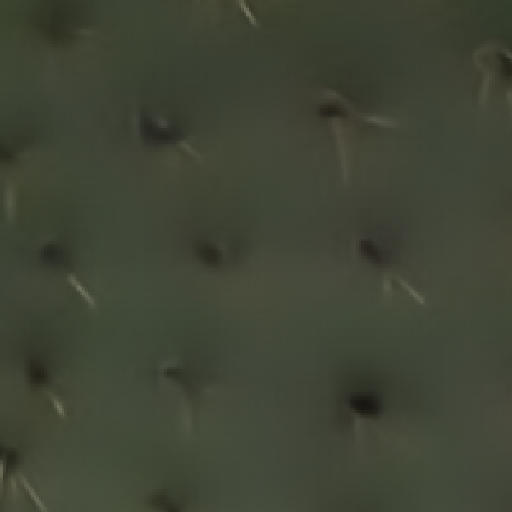}
    \end{subfigure}
    \begin{subfigure}[h]{0.21\linewidth}
    \centering
      \includegraphics[width=0.9\linewidth]{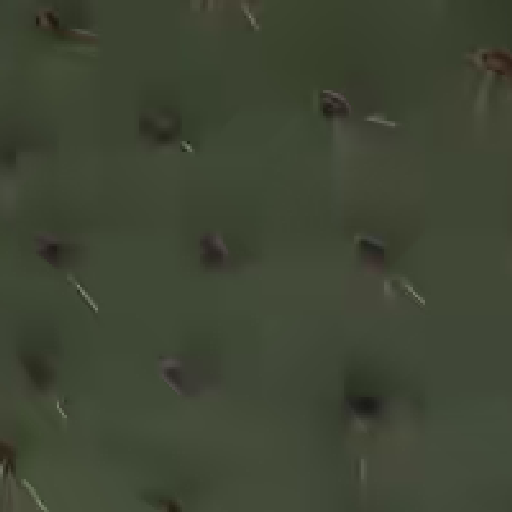}
    \end{subfigure}
    \begin{subfigure}[h]{0.21\linewidth}
    \centering
      \includegraphics[width=0.9\linewidth]{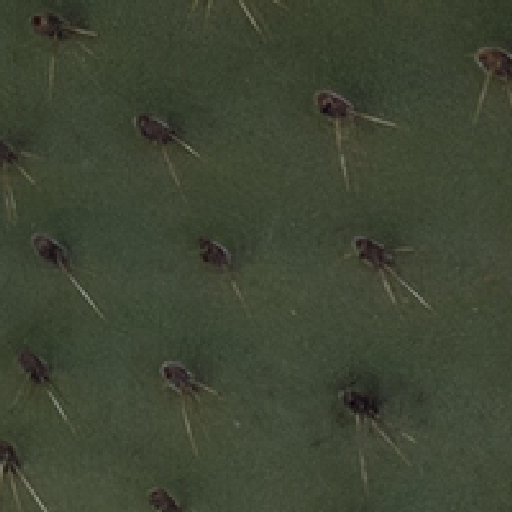}
    \end{subfigure}
    
    \vspace{1mm}

    \begin{subfigure}[h]{0.21\linewidth}
    \centering
    0.372 bpp
    \end{subfigure}
    \begin{subfigure}[h]{0.21\linewidth}
    \centering
    0.380 bpp
    \end{subfigure}
    \begin{subfigure}[h]{0.21\linewidth}
    \centering
    0.417 bpp
    \end{subfigure}
    \begin{subfigure}[h]{0.21\linewidth}
    \centering
    \BLK{GT}
    \end{subfigure}

    \vspace{1mm}

    \begin{subfigure}[h]{0.21\linewidth}
    \centering
      \caption{Ours}
    \end{subfigure}
    \begin{subfigure}[h]{0.21\linewidth}
    \centering
      \caption{ICLR18-Hyperprior}
    \end{subfigure}
    \begin{subfigure}[h]{0.21\linewidth}
    \centering
      \caption{BPG-4:4:4}
    \end{subfigure}
    \begin{subfigure}[h]{0.21\linewidth}
    \centering
      \caption{Ground Truth}
    \end{subfigure}

    \caption{Visualization of the reconstructed images by the proposed method, ICLR18-Hyperprior, and BPG.}
    \label{fig:visual} 
\end{figure*}

We conduct visual analysis on the reconstructed images in Fig.~\ref{fig:visual}, where we compare our method with the representative learning-based method (ICLR18-Hyperprior)~\cite{balle2018variational} and hybrid coding method (BPG-4:4:4)~\cite{bpg}. As shown, the proposed method reconstructs images with fewer artifacts at lower bit-rates. Specifically, the hybrid coding method relies on image partitioning. It inevitably causes blocking artifacts. Besides, the directional intra prediction scheme in the hybrid codec does not handle multi-directional edges well, as shown in Fig.~\ref{fig:visual}. More visual results are provided in the supplementary material.

\subsection{Cross-Metric Evaluation}
\label{sec:crossmetric}
\begin{figure*}[!h]
	\centering
\includegraphics[width=0.95\linewidth]{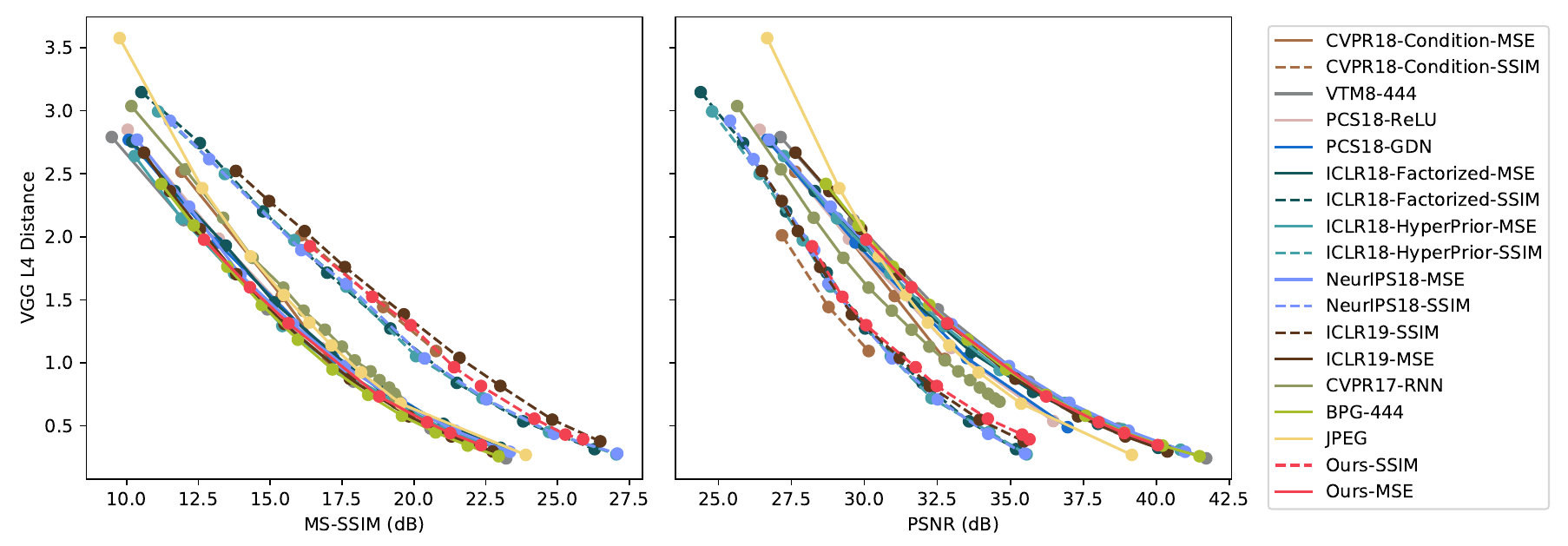}
\caption{
Evaluation of perceptual distance~\cite{johnson2016perceptual} (a lower value corresponds to better quality) with respect to PSNR and MS-SSIM for different methods. The methods correspond to those in Fig. \ref{fig:compare}.
}
	\label{fig:percept}
\end{figure*}

End-to-end learned image compression models can usually be trained towards many objectives as long as they are differentiable. Recent works usually evaluate two versions of the proposed method by training the model with both MSE (for PSNR) and MS-SSIM, as MS-SSIM better models visual quality for humans. Models trained on one objective may not perform well on the other metrics. Specifically, models trained with MS-SSIM as the loss function usually show lower PSNR values for a given range of bit-rates. Different models, which should achieve different performances, show similar levels of PSNR if they are trained using MS-SSIM. In contrast, different PSNR-optimized models do show different performance in MS-SSIM evaluation. For this class of models, those with higher results in PSNR usually perform better in MS-SSIM.

Because end-to-end learned models can be tuned with both PSNR and MS-SSIM, we are able to investigate the relationship between different metrics and objectives. We employ the perceptual metric, which is widely used in image enhancement and synthesis~\cite{johnson2016perceptual,zhang2018unreasonable}, as the metric for cross evaluation. Following the settings of perceptual loss in \cite{johnson2016perceptual}, the $L_2$ distance
of the output feature maps corresponding to four layers in the VGG-16~\cite{simonyan2014very} with respect to the original image and the reconstructed image are evaluated. Zhang \textit{et al.}~\cite{zhang2018unreasonable} show that the distance of the feature maps of such layers reflects the distortion with respect to human perception. Thus, we employ the metric as a supporting evaluation of the reconstruction quality for image compression methods.

To show the comparison, we plot the Perceptual-PSNR and Perceptual-MS-SSIM curves in Fig.~\ref{fig:percept}. Note that the distances with respect to the four layers are averaged for the illustration. Here are our observations of the experimental results.

\begin{itemize}
\item For a given level of PSNR, models trained on MS-SSIM show significantly less perceptual distortion, while for a given level of MS-SSIM, those trained on PSNR have less perceptual distortion. When a model is optimized for a metric, compared with others that are not optimized for that metric, the optimized one shows a higher perceptual distortion at the same level of the metric.
\item For models tuned with the MS-SSIM loss function, those with higher performance in MS-SSIM-bpp evaluation tend to result in larger perceptual distortion at a certain level of MS-SSIM. Although the same phenomenon is observed in the Perceptual-PSNR curve, it is not as significant.
\end{itemize}

To summarize, we observe in the experimental results that there exists a gap of different metrics, especially for models with better performance on one metric. Although end-to-end learning-based methods can be trained towards different objectives, they tend to be over-\textit{optimized} on that specific objective only. This phenomenon
is also related to
recent work on the investigation into the trade-off between perception and distortion~\cite{liu2019classification,blau2018perception}. In circumstances where we reserve more bit-rate for an image encoded with a better codec, as metrics such as PSNR and MS-SSIM show high enough values at that bit-rate, we may not be provided with the expected visual quality. A better assessment technique is needed, especially for the development of high-performance image compression methods. Furthermore, in real-world applications, the images are mostly consumed by human users, while there is a trend of developing image processing systems for machine vision tasks. To jointly optimize an image compression framework for both human perception and machine intelligence remains to be explored in future research.

\subsection{Discussion}

\begin{table}
	\centering
	\captionsetup{labelfont={color=black},font={color=black}}
	\caption{Encoding and decoding time (seconds) for various methods.}
	\label{tab:time}
	\begin{tabular}{cccccc}\toprule
		Task &Device &Ours & Context\cite{lee2018context} &VTM-8 \\\midrule
		\multirow{4}{*}{Encode} &1 Core &24.4 &50.1 &467.5 \\
		&2 Cores &13.6 &49.4 &467.5\footnotemark[4] \\
		&4 Cores &9.2 &48.4 &467.5\footnotemark[4] \\
		&4 Cores + GPU &4.9 &N/A &N/A \\
		\midrule
		\multirow{4}{*}{Decode} &1 Core &69.8 &186.5 &0.3 \\
		&2 Cores &41.4 &181.2 &0.3\footnotemark[4] \\
		&4 Cores &28.8 &177.7 &0.3\footnotemark[4] \\
		&4 Cores + GPU &7.1 &N/A &N/A \\
		\bottomrule
		\end{tabular}
\end{table}

\footnotetext[4]{We set QP=25 in this test. VTM-8 does not support multi-thread execution, so its time consumption should remain the same with different numbers of available cores.}

\subsubsection{Efficiency on Parallel Devices}
We also benchmark the encoding and decoding time for the proposed method, the context-model-based method~\cite{lee2018context} and VTM-8 as a hybrid codec. We test the encoder and decoder on a machine with Intel Core i7-7700K CPU and an NVIDIA RTX 2060 Super GPU. When doing encoding and decoding for an image from CLIC 19 validation dataset, we restrict the resources the program could utilize to test its time consumption on different parallel scales. The results are presented in Table~\ref{tab:time}. As shown, while the proposed method achieves competitive rate-distortion performance against context-model based methods, as it does not rely on the context model, it is not limited to using a serial decoding scheme and therefore runs faster than those methods in the experiments. VTM is designed with no thread-level parallelism, thus not accelerated by multi-core devices.

\subsubsection{Comparison with VVC}
We compare the learning-based image compression methods with VVC~\cite{vtm8}, the advanced hybrid transform coding scheme. Based on the experimental results, we summarize different characteristics of VVC and the advanced learning-based methods.

\textbf{VVC.} VVC has improved design in picture partitioning, intra prediction, transform, and quantization \textit{etc}. over HEVC~/~BPG, and it achieves better rate-distortion performance than benchmarked end-to-end learned image compression methods. Besides, as the rate-distortion optimization (RDO) is done only at the encoder side, VVC decoder has significantly lower complexity. Thus, it better meets the need in most real-world applications.

\textbf{Learning-Based.} Advanced learned image compression methods adopt neural networks to learn the image encoder and decoder automatically. Thanks to the rapid evolution of machine learning techniques, tremendous improvements in rate-distortion performance have been witnessed in the past five years. These methods tend to be more flexible than hand-crafted hybrid coding methods, as they can be end-to-end optimized to avoid conflicts between components. They can also be potentially accelerated by parallel computing devices. Existing works have shown the potentials of end-to-end learned to achieve higher performance and better efficiency in the near future.

\section{Conclusion}
\label{sec:conclude}

In this paper, we conduct a systematic benchmark on existing methods for learned image compression. We first summarize the contributions of existing works, with novelties highlighted, and we also analyze and discuss insights and challenges in this problem. With inspiration from the technical merits, we propose a coarse-to-fine hyperprior framework for image compression, trying to address the issues of existing methods in multiresolution context modeling. We conduct a thorough evaluation of existing methods and the proposed method, which illustrates the great progress made in the research, as well as the driving force for such advancements. The results also demonstrate the superiority of the proposed method in handling images with various content and resolutions. Further cross-metric evaluation indicates the future research direction of jointly optimizing an image compression method for both machine intelligence systems and human perception.

\section*{Acknowledgments}

This work is supported by the National Key Research and Development Program of China under Grant No.~2018AAA0102702, the Fundamental Research Funds for the Central Universities, and the National Natural Science Foundation of China under Contract No.~61772043 and No.~62022038.

{\footnotesize
\bibliographystyle{IEEEtran}
\bibliography{egbib}
}

\begin{IEEEbiography}[{\includegraphics[width=1in,height=1.25in,keepaspectratio]{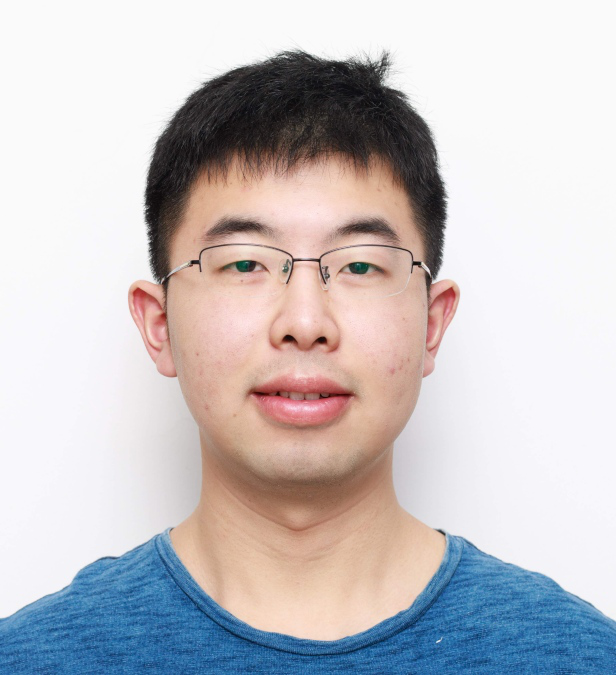}}]
{Yueyu Hu} (STM'18-GSM'19)
received the B.S. degree in computer science from Peking University, Beijing, China, in 2018, where he is currently working toward the master's degree with Wangxuan Institute of Computer Technology, Peking University. His current research interests include video and image compression and analytics with machine learning.
\end{IEEEbiography}
\vfill
\begin{IEEEbiography}[{\includegraphics[width=1in,height=1.25in,keepaspectratio]{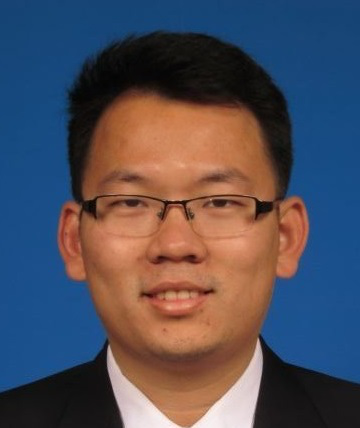}}]
	{Wenhan Yang} (M'18) 
	received the B.S degree and Ph.D. degree (Hons.) in computer science from Peking University, Beijing, China, in 2012 and 2018. He is currently a postdoctoral research fellow with the Department of Computer Science, City University of Hong Kong. Dr. His current research interests include image/video processing/restoration, bad weather restoration, human-machine collaborative coding. He has authored over 100 technical articles in refereed journals and proceedings, and holds 9 granted patents.
	He received the IEEE ICME-2020 Best Paper Award, the IFTC 2017 Best Paper Award, and the IEEE CVPR-2018 UG2 Challenge First Runner-up Award.
	He was the Candidate of CSIG Best Doctoral Dissertation Award in 2019. He served as the Area Chair of IEEE ICME-2021, and the Organizer of IEEE CVPR-2019/2020/2021 UG2+ Challenge and Workshop.
\end{IEEEbiography}
\vfill
\begin{IEEEbiography}[{\includegraphics[width=1in,height=1.25in,keepaspectratio]{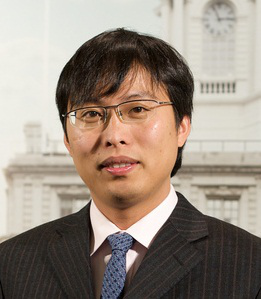}}]
{Zhan Ma} (SM'19) received the B.S. and M.S. from Huazhong University of Science and Technology (HUST), Wuhan, China, in 2004 and 2006 respectively, and the Ph.D. degree from the New York University, New York, in 2011. He is now on the faculty of Electronic Science and Engineering School, Nanjing University, Jiangsu, 210093, China. From 2011 to 2014, he has been with Samsung Research America, Dallas TX, and Futurewei Technologies, Inc., Santa Clara, CA, respectively. His current research focuses on the next-generation video coding, energy-efficient communication, gigapixel streaming and deep learning. He is a co-recipient of 2018 ACM SIGCOMM Student Research Competition Finalist, 2018 PCM Best Paper Finalist, and 2019 IEEE Broadcast Technology Society Best Paper Award.
\end{IEEEbiography}
\vfill
\begin{IEEEbiography}[{\includegraphics[width=1in,height=1.25in,clip,keepaspectratio]{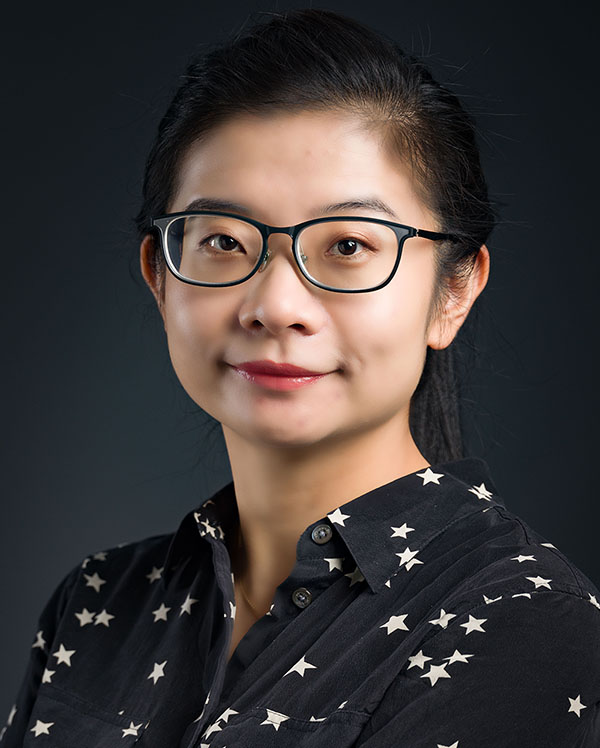}}]
	{Jiaying Liu} (M'10-SM'17) 
	is currently an Associate Professor, Peking University Boya Young Fellow with the Wangxuan Institute of Computer Technology, Peking University. She received the Ph.D. degree (Hons.) in computer science from Peking University, Beijing China, 2010. She has authored over 100 technical articles in refereed journals and proceedings, and holds 50 granted patents. Her current research interests include multimedia signal processing, compression, and computer vision. 
	
	Dr. Liu is a Senior Member of IEEE, CSIG and CCF. She was a Visiting Scholar with the University of Southern California, Los Angeles, from 2007 to 2008. She was a Visiting Researcher with the Microsoft Research Asia in 2015 supported by the Star Track Young Faculties Award. She has served as a member of Multimedia Systems \& Applications Technical Committee (MSA TC), and Visual Signal Processing and Communications Technical Committee (VSPC TC) in IEEE Circuits and Systems Society. She received the IEEE ICME-2020 Best Paper Award and IEEE MMSP-2015 Top10\% Paper Award. She has also served as the Associate Editor of IEEE Trans. on Image Processing, IEEE Trans. on Circuit System for Video Technology and Elsevier JVCI, the Technical Program Chair of IEEE ICME-2021/ACM ICMR-2021, the Publicity Chair of IEEE ICME-2020/ICIP-2019, and the Area Chair of CVPR-2021/ECCV-2020/ICCV-2019. She was the APSIPA Distinguished Lecturer (2016-2017).
\end{IEEEbiography}
\vfill
\end{document}